\documentclass[a4paper,twocolumn,11pt,accepted=2020-05-12]{quantumarticle}
\pdfoutput=1
\usepackage[utf8]{inputenc}
\usepackage[english]{babel}
\usepackage[T1]{fontenc}
\usepackage{amsmath}
\usepackage{hyperref}

\usepackage{tikz}
\usepackage{lipsum}

\usepackage{multirow}
\usepackage{amssymb,amstext,bm,bbm}
\newcommand{\red}[1]{#1}
\newcommand{\blue}[1]{#1}
\newcommand{\purple}[1]{#1}
\usepackage{tikz-cd}

\newcommand{\bra}[1]{{\langle{#1}\vert}}
\newcommand{\ket}[1]{{\vert{#1}\rangle}}
\newcommand{\bbra}[1]{{\langle\!\langle{#1}\vert}}
\newcommand{\kket}[1]{{\vert{#1}\rangle\!\rangle}}
\newcommand{\bracket}[2]{\langle #1 \vert #2 \rangle}

\newcommand{\komut}[2]{ \, [\, #1\, ,\, #2\, ]\, }
\newcommand{\akomut}[2]{\, \{\, #1\, ,\, #2\, \}\, }

\newcommand{\tr}{\mathop{\rm Tr}\nolimits}
\renewcommand{\Re}{\mathop{\rm Re}\nolimits}

\newcommand{\prirodni}{\ensuremath{\mathbb{N}}}
\newcommand{\realni}{\ensuremath{\mathbb{R}}}
\newcommand{\kompleksni}{\ensuremath{\mathbb{C}}}

\newcommand{\sspan}{\mathop{\rm span}\nolimits}

\newcommand{\one}{\mathbbm{1}}

\newcommand{\ds}{\displaystyle}

\newcommand{\cC}{{\cal C}}

\newcommand{\cF}{{\cal F}}
\newcommand{\cG}{{\cal G}}
\newcommand{\cH}{{\cal H}}
\newcommand{\cI}{{\cal I}}

\newcommand{\cL}{{\cal L}}
\newcommand{\cM}{{\cal M}}

\newcommand{\cO}{{\cal O}}
\newcommand{\cP}{{\cal P}}

\newcommand{\orderi}{\prec_{I}}
\newcommand{\orderc}{\prec_{\cC}}
\newcommand{\orderm}{\prec_{\cM}}

\begin{document}

\title{Causal orders, quantum circuits and spacetime: distinguishing between definite and superposed causal orders}

\author{Nikola Paunkovi\'c}
\email{npaunkov@math.tecnico.ulisboa.pt}
\orcid{0000-0002-9345-4321}
\affiliation{Instituto de Telecomunica\c{c}\~oes and Departamento de Matem\'atica, Instituto Superior T\'ecnico, Universidade de Lisboa, Avenida Rovisco Pais 1049-001, Lisboa, Portugal}

\author{Marko Vojinovi\'c}
\email{vmarko@ipb.ac.rs}
\orcid{0000-0001-6977-4870}
\affiliation{Institute of Physics, University of Belgrade, Pregrevica 118, 11080 Belgrade, Serbia}

\begin{abstract}
We study the notion of causal orders for the cases of (classical and quantum) circuits and spacetime events. We show that every circuit can be immersed into a classical spacetime, preserving the compatibility between the two causal structures. Using the process matrix formalism, we analyse the realisations of the quantum switch using 4 and 3 spacetime events in classical spacetimes with fixed causal orders, and the realisation of a gravitational switch with only 2 spacetime events that features superpositions of different gravitational field configurations and their respective causal orders. We show that the current quantum switch experimental implementations do not feature superpositions of causal orders between spacetime events, and that these superpositions can only occur in the context of superposed gravitational fields. We also discuss a recently introduced operational notion of an event, which does allow for superpositions of respective causal orders in flat spacetime quantum switch implementations. We construct two  observables that can distinguish between the quantum switch realisations in classical spacetimes, and gravitational switch implementations in superposed spacetimes. Finally, we discuss our results in the light of the modern relational approach to physics.
\end{abstract}

\maketitle

\section{\label{sec:intro}Introduction}

The notion of causality is one of the most prominent in science, and also in philosophy of Nature. Its treatment separates Aristotelian from the modern physics, and its {\em formal} meaning within the latter is likely to have played a significant role, over the past centuries since Galileo, in forming our current {\em everyday} understanding of the notion of causality. While in Newtonian physics the cause-effect relations were encompassed by a rather simple linear and absolute time, Einstein's analysis of causal relations was pivotal in the formulation of the theory of relativity. But it was quantum mechanics (QM) that, through the EPR argument~\cite{ein:pod:ros:35}, further formalised by Bell~\cite{bel:64}, showed how quantum nonlocality, rooted in the superposition principle of QM, revolutionised our everyday notion of causality. Finally, strong theoretical evidence that, when combining the two fundamental theories of the modern physics, one is to expect explicit dynamical nonlocal effects in quantum gravity (QG), shows that our basic understanding of causality and causal orders might be crucial in the development of new physics.

Recently, causal orders were, mainly within the quantum information community, discussed in the context of controlled operations. In particular, it was argued that the quantum switch, a specific controlled operation introduced in~\cite{chi:dar:per:val:13}, exhibits superpositions of causal orders, not only in the context of quantised gravity, where genuine superpositions of different states of gravity are present, but also in the experimental realisations performed in classical spacetimes with fixed causal structure~\cite{pro:etal:15,rub:roz:fei:ara:zeu:pro:bru:wal:17,rub:roz:mas:ara:zyc:bru:wal:17}.
\blue{Note that the notion of causal order discussed in these papers is {\em different} from the causal order of the underlying spacetime structure. We discuss in detail the relation between the two.}

In this paper, we analyse the notion of causal orders in the context of classical and quantum circuits\blue{, and relate it to the spacetime causal structures}. We prove that each circuit can be realised in a classical spacetime, preserving \blue{the} fixed causal relations \blue{of the former, with respect to the causal relations between spacetime events of the latter} (see the next section for the details of the theorem). Further, we analyse possible realisations of the quantum switch, showing that those performed in everyday labs do not feature superpositions of causal orders \blue{between spacetime events} (consistent with our theorem), but rather standard non-relativistic quantum mechanical (coherent) superpositions of different evolutions of a system. On the other hand, we argue that genuine superpositions of different causal orders are indeed to be expected within the QG scenario, where superpositions of different states of the gravitational field, with their corresponding causal orders, are manifestly allowed (Hardy was one of the first to discuss the notion of superpositions of causal orders in the context of QG~\cite{har:07}). \blue{In addition, we explicitly construct two distinct observables that can distinguish between the realisations of the quantum switch in classical spacetimes, and implementations of the gravitational switch in superposed spacetimes. This way, we show that the two notions of causal orders, namely one discussed in ~\cite{pro:etal:15,rub:roz:fei:ara:zeu:pro:bru:wal:17,rub:roz:mas:ara:zyc:bru:wal:17} and the other discussed in this paper, can be experimentally distinguished,} in contrast to the opposite claim present in the literature~\cite{pro:etal:15}. Finally, we discuss our results in the context of the relational approach to physics.

The layout of the paper is as follows. In Section \ref{sec:causal_orders}, we introduce the notion of causal order for circuits, and prove the Theorem of the circuit immersion in classical spacetimes. Section \ref{sec:quantum_switch} is devoted to the analysis of the quantum switch implementations in classical spacetimes that do not feature superpositions of \blue{spacetime} causal orders, as well as implementations in the context of QG. In Section~\ref{sec:distinguishing}, we compare the quantum switch implementations discussed, and introduce observables that can distinguish between those that feature superpositions of \blue{spacetime} causal orders, and those that do not. Section~\ref{sec:discussion} is devoted to the discussion of the superpositions of causal orders in the context of the relational approach to physics. Finally, in Section~\ref{sec:conclusions}, we present and discuss the results, provide some final remarks, and list possible future research directions.

\section{\label{sec:causal_orders}Causal orders}

We begin by discussing circuits and their realisations in (classical) spacetimes with well defined fixed causal orders. Given a directional acyclic graph $G = (I,E)$, where $I$ is the set of graph nodes, and $E = \{ (u,v)\, |\, u,v \in I\}$ is the set of its directed edges (arrows pointing from $u$ to $v$ representing the {\em wires} of the circuit), a {\em circuit} $\cC$ over the set of operations $\cG$ is a pair $\cC = (G,g)$, where the mapping $g: I \rightarrow \cG$ assigns operations to each node. Depending on the type of the operations from $\cG$, we will call the circuit {\em classical} (if the operations are, say, classical logic gates), or {\em quantum} (if the operations are, say, unitaries, measurements, etc.).

The fact that $G$ is directional and acyclic allows one to define a {\em partial order} $\orderi$ over the set $I$ as
$$
u \orderi v \quad \stackrel{\text{def}}{\Longleftrightarrow} \hphantom{mmmmmmmm}
$$
$$
\Big(\exists \ n \in \mathbb N \ \wedge \  \{u \equiv u_1, u_2, \dots, u_n \equiv v\} \subset I\Big)
$$
\begin{equation}
\Big(\forall \ i \in \{ 1,2, \dots, n\!-\!1\}\Big) \ (u_i,u_{i+1})\in E\,,
\end{equation}
representing the causal relation between the graph nodes. Next, we define the set of {\em gates of the circuit} $\cC$ as $\cG_\cC = \{g_u \equiv (u,g(u)) \, |\, u \in I\}$. The induced causal order between the circuit gates $\orderc$ is by definition given~as
\begin{equation} \label{eq:giOrderEquivalence}
g_u \orderc g_v \quad \stackrel{\text{def}}{\Longleftrightarrow} \quad u \orderi v\,.
\end{equation}
Moreover, since there exists a canonical bijection between $I$ and $\cG_\cC$, the order relations $\orderi$ and $\orderc$ are {\em isomorphic}.

Finally, we can introduce the set $\cM$ of all spacetime events, which is assumed to be a traditional $4D$ manifold. On this spacetime manifold we assume to have a gravitational field, described in a standard way, using a metric tensor $g_{\mu\nu}$. The metric is assumed to be of Minkowski signature, such that the metric-induced light cone structure determines a partial order relation between nearby events, denoted~$\orderm^g$ (or simply $\orderm$ when the choice of the metric is implicit). Note that the causal order over the spacetime events is not an intrinsic property of the spacetime manifold itself, but rather determined by the metric, i.e., the configuration of the gravitational field living on the manifold.

One might pose a question if, given a formal circuit $\cC$ with gates $\cG_\cC$, it is possible to realise it in a lab --- if it is possible to ``immerse'' it into spacetime. More precisely, given an arbitrary spacetime manifold $\cM$, our goal is to study if there exists an {\em order-preserving map} $ \cP: \cG_\cC \to \cM$, i.e., if the partial order relations satisfy
\begin{equation} \label{eq:ImmersingRequirement}
g_u \orderc g_v \;\; \Longrightarrow \;\; \cP(g_u) \orderm \cP(g_v)\,,
\end{equation}
for every $g_u,g_v \in \cG_\cC$. To that end, we formulate the following theorem (the proof is given in Appendix \ref{sec:AppProof}).

\medskip

\textbf{Theorem.} \textit{Any circuit $\cC$ can be immersed into a globally hyperbolic spacetime manifold $\cM$, such that its relation of partial order $\orderc$ is preserved by the relation of spacetime events $\orderm$.}

\medskip

Regarding the physical interpretation of the Theorem, note that it assigns a spacetime point to each gate in a circuit, as opposed to a point in $3D$ space. Since each spatially localised apparatus may perform the same operation more than once, at different moments in time, it may then correspond to several different gates of the circuit, and thus several different nodes of the graph, instead of just one. In other words, a single piece of experimental equipment {\em does not} always correspond to a single gate of a circuit.

In addition to the above comment, note that in reality each operation actually takes place in some finite volume of both space and time. However, in theoretical arguments it is convenient to approximate this finite spacetime volume with a single point, ignoring the size and time of activity of the device performing the operation. We adopt this approximation throughout this paper.

Circuits are seen as operations acting upon certain inputs to obtain the corresponding outputs. Usually, the initial/final states (which include instructions, measurement results, etc.) are depicted by the wires. But in our approach, the input state is prepared by the ``initial gate'' $\cI$, while the output state is obtained by the ``final gate'' $\cF$. This way, the circuit $\cC$ is seen as an operation $\cO_\cC$ acting from $\cI$ to $\cF$.

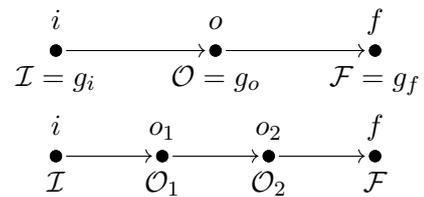
\begin{figure}[h!]
\begin{center}
\begin{tikzpicture}[scale=0.7]
\node at (0,2) (upSi) {};
\node at (3,2) (upO) {};
\node at (6,2) (upSf) {};
\node at (0,0) (downSi) {};
\node at (2,0) (downOone) {};
\node at (4,0) (downOtwo) {};
\node at (6,0) (downSf) {};
\filldraw[black] (upSi) circle (3pt) node[anchor=south] {$i\strut$};
\filldraw[black] (upSi) circle (3pt) node[anchor=north] {$\mathcal{I}=g_i\strut$};
\filldraw[black] (upO) circle (3pt) node[anchor=south] {$o\strut$};
\filldraw[black] (upO) circle (3pt) node[anchor=north] {$\cO=g_o\strut$};
\filldraw[black] (upSf) circle (3pt) node[anchor=south] {$f\strut$};
\filldraw[black] (upSf) circle (3pt) node[anchor=north] {$\mathcal{F}=g_f\strut$};
\filldraw[black] (downSi) circle (3pt) node[anchor=south] {$i\strut$};
\filldraw[black] (downSi) circle (3pt) node[anchor=north] {$\mathcal{I}\strut$};
\filldraw[black] (downOone) circle (3pt) node[anchor=south] {$o_1\strut$};
\filldraw[black] (downOone) circle (3pt) node[anchor=north] {$\cO_1\strut$};
\filldraw[black] (downOtwo) circle (3pt) node[anchor=south] {$o_2\strut$};
\filldraw[black] (downOtwo) circle (3pt) node[anchor=north] {$\cO_2\strut$};
\filldraw[black] (downSf) circle (3pt) node[anchor=south] {$f\strut$};
\filldraw[black] (downSf) circle (3pt) node[anchor=north] {$\mathcal{F}\strut$};
\draw[->] (upSi) -- (upO);
\draw[->] (upO) -- (upSf);
\draw[->] (downSi) -- (downOone);
\draw[->] (downOone) -- (downOtwo);
\draw[->] (downOtwo) -- (downSf);
\end{tikzpicture}
\end{center}
\caption{\blue{Implementing operation $\cO$ with a single gate (upper diagram), and by two consecutive gates $\cO_1$ and $\cO_2$ (lower diagram).}}
\label{sl:jedan}
\end{figure}

Note that, given a circuit $\cC$, the corresponding overall operation $\cO_\cC$ (as well as the input and the output gates $\mathcal{I}$ and $\mathcal{F}$) is uniquely defined. The opposite is not the case: given the operation $\cO$, one can design different circuits $\cC, \cC^\prime, \dots$  that achieve it. To see this, let us consider the simplest case of the operation which satisfies $\cO = \cO_2 \circ \cO_1$, where $\circ$ represents the composition of operations. This operation can be trivially achieved by the two circuits: (i) $\cC$, which consists of three nodes --- node $i$  whose gate $\mathcal{I}$ prepares the input state, node $o$ that applies the gate $g_o = \cO$, and node $f$ whose gate $\mathcal{F}$ outputs either the quantum state, the classical outcome(s), or the combination of the two; (ii) $\cC_{12}$, which consists of four nodes --- nodes $i$ and $f$ that perform the same operations as before, and {\em two} intermediate nodes $o_1$ and $o_2$ that perform $g_{o_1} = \cO_1$ and $g_{o_2} = \cO_2$, respectively. \blue{For simplicity, here and elsewhere in the text, by $\cO$ we denote both the operation and the gate that implements it.} The two situations are depicted in the following diagrams \blue{(see Figure~\ref{sl:jedan})}.

\blue{Finally, in recent literature one can find a notion of an event which is different from the notion of a spacetime point~\cite{chi:dar:per:val:13,pro:etal:15,rub:roz:fei:ara:zeu:pro:bru:wal:17,rub:roz:mas:ara:zyc:bru:wal:17,ore:cos:bru:12,ara:bra:cos:fei:gia:bru:15,bad:ara:bru:que:19,gue:rub:bru:18}. Namely, one can talk about events as interactions between the quantum system under consideration and the apparatus in the lab. This is motivated by the operational approach to physics, where the interactions between objects are taken as fundamental. Then, one can introduce the relation of partial order, which reflects the causal relationships between such events. Of course, in general, this causal order does not need to coincide with the spacetime causal order. Throughout this paper, if not explicitly stated otherwise, by causal order we mean the order between the spacetime points, which due to our Theorem can also be regarded as the order between the circuit gates. We discuss the difference between the two notions of causal orders in Section~\ref{sec:distinguishing}.}

\section{\label{sec:quantum_switch} Quantum switch}

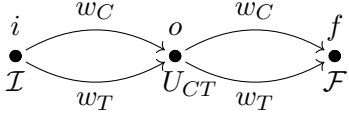
\begin{figure}[h!]
\begin{center}
\begin{tikzpicture}[scale=0.7]
\node at (-1,0) (inode) {};
\node at (2,0) (onode) {};
\node at (5,0) (fnode) {};

\filldraw[black] (inode) circle (3pt) node[anchor=south] {$i\strut$};
\filldraw[black] (inode) circle (3pt) node[anchor=north] {$\mathcal{I}\strut$};
\filldraw[black] (onode) circle (3pt) node[anchor=south] {$o\strut$};
\filldraw[black] (onode) circle (3pt) node[anchor=north] {$\quad U_{CT}\strut$};
\filldraw[black] (fnode) circle (3pt) node[anchor=south] {$f\strut$};
\filldraw[black] (fnode) circle (3pt) node[anchor=north] {$\mathcal{F}\strut$};

\draw[->] (inode) to [out=30,in=150] (onode);
\draw[->] (inode) to [out=-30,in=-150] (onode);
\draw[->] (onode) to [out=30,in=150] (fnode);
\draw[->] (onode) to [out=-30,in=-150] (fnode);

\filldraw[black] (0.5,0.5) circle (0pt) node[anchor=south] {$w_C$};
\filldraw[black] (0.5,-0.5) circle (0pt) node[anchor=north] {$w_T$};
\filldraw[black] (3.5,0.5) circle (0pt) node[anchor=south] {$w_C$};
\filldraw[black] (3.5,-0.5) circle (0pt) node[anchor=north] {$w_T$};

\end{tikzpicture}
\end{center}
\caption{\blue{Controlled operation $U_{CT}$. Applying operation $U_b$ on a system in the wire $w_T$ {\em controlled} by the state $\ket{b}$ on a system in the wire $w_C$, with $b=0,1$.}}
\label{sl:dva}
\end{figure}

The most prominent feature of quantum systems is that they can be found in {\em coherent superpositions} of states. This allows for applying the so-called {\em control operations}. For simplicity, let us assume that operations $\cO$ are unitaries, denoted as $U$. Given a {\em control} system $C$ in a superposition $\ket{\varphi}_C = a \ket{0}_C + b \ket{1}_C$ (with $\bracket{0}{1}_C = 0$), the control operation 
\begin{equation}
\label{eq:control_operation}
U_{CT} = \ket{0}_C\bra{0} \otimes U_0 + \ket{1}_C\bra{1} \otimes U_1
\end{equation}
transforms the initial product state $\ket{\Psi_i}_{CT} = \ket{\varphi}_C \otimes \ket{\psi_i}_T$  between the control and the {\em target} systems into the final entangled state $\ket{\Psi_f}_{CT} = a \ket{0}_C \otimes U_0 \ket{\psi_i}_T + b \ket{1}_C \otimes U_1 \ket{\psi_i}_T$. A simple realisation of such operation by a circuit consisting of three gates is shown below \blue{(see Figure~\ref{sl:dva})}.

Here, the first node and the corresponding gate prepares the initial superposition of the control system, the second implements $U_{CT}$, and the third is either an identity, a measurement on the two systems, or a combination (say, a measurement of the target qubit, while leaving the control intact). In order to allow for the description of quantum superpositions, we introduce the notion of a vacuum in the analysis of quantum circuits, as is done for example in \cite{por:mat:mau:ren:tac} (for technical details, see Appendix \ref{sec:AppI}).

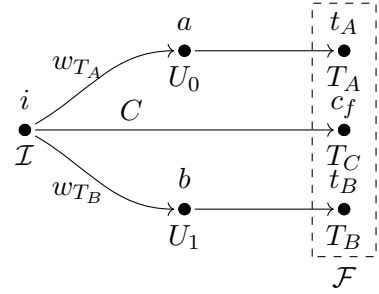
\begin{figure}[h!]
\begin{center}
\begin{tikzpicture}[scale=0.7]
\node at (-1,0) (inode) {};
\node at (2,1.5) (anode) {};
\node at (2,-1.5) (bnode) {};

\node at (5,1.5) (tanode) {};
\node at (5,-1.5) (tbnode) {};
\node at (5,0) (cfnode) {};

\filldraw[black] (inode) circle (3pt) node[anchor=south] {$i\strut$};
\filldraw[black] (inode) circle (3pt) node[anchor=north] {$\mathcal{I}\strut$};
\filldraw[black] (anode) circle (3pt) node[anchor=south] {$a\strut$};
\filldraw[black] (anode) circle (3pt) node[anchor=north] {$U_0\strut$};
\filldraw[black] (bnode) circle (3pt) node[anchor=south] {$b\strut$};
\filldraw[black] (bnode) circle (3pt) node[anchor=north] {$U_1\strut$};
\filldraw[black] (tanode) circle (3pt) node[anchor=south] {$t_A\strut$};
\filldraw[black] (tanode) circle (3pt) node[anchor=north] {$T_A\strut$};
\filldraw[black] (tbnode) circle (3pt) node[anchor=south] {$t_B\strut$};
\filldraw[black] (tbnode) circle (3pt) node[anchor=north] {$T_B\strut$};
\filldraw[black] (cfnode) circle (3pt) node[anchor=south] {$c_f\strut$};
\filldraw[black] (cfnode) circle (3pt) node[anchor=north] {$T_C\strut$};

\draw[->] (inode) to [out=30,in=180] (anode);
\draw[->] (inode) to [out=-30,in=180] (bnode);
\draw[->] (inode) to (cfnode);
\draw[->] (anode) to (tanode);
\draw[->] (bnode) to (tbnode);

\filldraw[black] (0,0.8) circle (0pt) node[anchor=south] {$w_{T_A}$};
\filldraw[black] (0,-0.8) circle (0pt) node[anchor=north] {$w_{T_B}$};
\filldraw[black] (1,0) circle (0pt) node[anchor=south] {$C$};

\draw[dashed] (4.4,-2.4) rectangle (5.6,2.4);
\node[anchor=north] at (5,-2.4) {$\mathcal{F}$};

\end{tikzpicture}
\end{center}
\caption{\blue{Implementation of the controlled operation using the spatial degree of freedom as a control.}}
\label{sl:tri}
\end{figure}

As noted above, given the operation, many different circuits can achieve it. Indeed, in standard optical implementations of the above controlled operation~\eqref{eq:control_operation}, the control qubit is spanned by two spatial modes of a photon, while the target one is its polarisation degree of freedom. The initial superposition state of the control qubit is prepared by a beam splitter, while the two operations $U_0$ and $U_1$ are implemented locally in Alice's and Bob's laboratories. Note that, since the control qubit is achieved by the means of two spatial modes of a single photon, while the target qubit is, being the photon's polarisation, ``attached to'' the control, the target is formally achieved by two degrees of freedom (two wires), one assigned to Alice ($T_A$), and the other to Bob ($T_B$). Thus, in such a realization, the control degree of freedom is redundant in the circuit diagram and can be omitted. Nevertheless, since we will later discuss the case of the gravitational quantum switch, in which the gravitational degree of freedom plays the role of the control, here we keep its corresponding wire and gate in the diagram, as presented below \blue{(see Figure~\ref{sl:tri})}.

The final gate $\mathcal{F}$ consists of three ``elementary gates'', represented by the circuit nodes $t_A$ and $t_B$ for the two target wires, and the node $c_f$ for the final control wire.

\begin{figure}[h!]
\begin{center}
\begin{tikzpicture}[scale=0.7]
\node at (-1,0) (inode) {};
\node at (2,1.5) (aunode) {};
\node at (2,-1.5) (bvnode) {};
\node at (3.5,1.5) (avnode) {};
\node at (3.5,-1.5) (bunode) {};
\node at (5,1.5) (tanode) {};
\node at (5,-1.5) (tbnode) {};
\node at (5,0) (cfnode) {};

\filldraw[black] (inode) circle (3pt) node[anchor=south] {$i\strut$};
\filldraw[black] (inode) circle (3pt) node[anchor=north] {$\mathcal{I}\strut$};
\filldraw[black] (aunode) circle (3pt) node[anchor=south] {$a_U\strut$};
\filldraw[black] (aunode) circle (3pt) node[anchor=north] {$U\strut$};
\filldraw[black] (bvnode) circle (3pt) node[anchor=south] {$b_V\strut$};
\filldraw[black] (bvnode) circle (3pt) node[anchor=north] {$V\strut$};
\filldraw[black] (avnode) circle (3pt) node[anchor=south] {$a_V\strut$};
\filldraw[black] (avnode) circle (3pt) node[anchor=north] {$V\strut$};
\filldraw[black] (bunode) circle (3pt) node[anchor=south] {$b_U\strut$};
\filldraw[black] (bunode) circle (3pt) node[anchor=north] {$U\strut$};
\filldraw[black] (tanode) circle (3pt) node[anchor=south] {$t_A\strut$};
\filldraw[black] (tanode) circle (3pt) node[anchor=north] {$T_A\strut$};
\filldraw[black] (tbnode) circle (3pt) node[anchor=south] {$t_B\strut$};
\filldraw[black] (tbnode) circle (3pt) node[anchor=north] {$T_B\strut$};
\filldraw[black] (cfnode) circle (3pt) node[anchor=south] {$c_f\strut$};
\filldraw[black] (cfnode) circle (3pt) node[anchor=north] {$T_C\strut$};

\draw[->] (inode) to [out=30,in=180] (aunode);
\draw[->] (inode) to [out=-30,in=180] (bvnode);
\draw[->] (aunode) to (avnode);
\draw[->] (bvnode) to (bunode);
\draw[->] (inode) to (cfnode);
\draw[->] (avnode) to (tanode);
\draw[->] (bunode) to (tbnode);

\filldraw[black] (0,0.8) circle (0pt) node[anchor=south] {$w_{T_A}$};
\filldraw[black] (0,-0.8) circle (0pt) node[anchor=north] {$w_{T_B}$};
\filldraw[black] (1,0) circle (0pt) node[anchor=south] {$C$};

\draw[dashed] (4.4,-2.4) rectangle (5.6,2.4);
\node[anchor=north] at (5,-2.4) {$\mathcal{F}$};

\end{tikzpicture}
\end{center}
\caption{\blue{The quantum switch.}}
\label{sl:cetiri}
\end{figure}
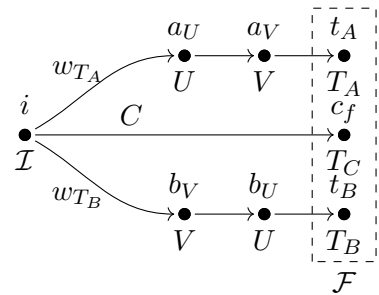

An important instance of controlled operations is the so-called {\em quantum switch}, for which the two controlled operations are given by $U_0 = UV$ and $U_1 = VU$, where $U$ and $V$ are two arbitrary unitaries~\cite{chi:dar:per:val:13}. Having two pairs of equipment, one applying $U$ and the other $V$, it is straightforward to implement the quantum switch through the circuit similar to the one above, which instead of two gates, one in the node $a$ applying $U_0$, and another in node $b$ applying $U_1$, contains four gates placed in the nodes $a_U$, $a_V$, $b_V$ and $b_U$ \blue{(see Figure~\ref{sl:cetiri})}.

The question arises, is it possible to achieve the same using less resources, say, using only two such pieces of equipment, located in two different points (regions) of 3D {\em space}? Indeed, it is possible to do so, and recently a number of implementations of the quantum switch were performed {in flat Minkowski spacetime}~\cite{pro:etal:15,rub:roz:fei:ara:zeu:pro:bru:wal:17,rub:roz:mas:ara:zyc:bru:wal:17}. Nevertheless, such implementations still correspond to circuits that implement $U_0$ and $U_1$ by four, rather than two gates. The difference is that, when immersing it in a flat spacetime, the two pairs of gates are now distinguished only by the temporal, rather than all four spacetime coordinates. Thus, one cannot talk of superpositions of causal orders \blue{between spacetime events} in such implementations, as flat (indeed, any globally hyperbolic) spacetime has a manifestly fixed causal order. To implement $U_0$ and $U_1$ of the quantum switch by a circuit that consists of two gates only (and thus two corresponding spacetime points), one needs a superposition of gravitational fields with different (incompatible) causal orders. In the following two subsections, we analyse in more detail the ``4-event'' and the ``3-event'' implementations of the quantum switch, while the ``2-event'' case is discussed in the last subsection (the numbers 4, 3 and 2 refer to the numbers of \blue{spacetime} events corresponding to distinct gates used to achieve $U_0$ and $U_1$). A detailed mathematical description using the process matrix formalism~\cite{ore:cos:bru:12}, is presented in the Appendices~\ref{sec:AppFourEvent},~\ref{sec:AppThreeEvent} and~\ref{sec:AppTwoEvent}.

\blue{Following the previously mentioned distinction between the spacetime event and the operational notion of the event, the $4$-event and $3$-event quantum switch implementations will have a description within the operational approach that is different from the spacetime description. In particular, in such approach these two implementations of quantum switch would feature only $2$ operationally defined events, and thus the superposition of the corresponding causal orders.}

\subsection{4-event process}
\label{sec:4-event}

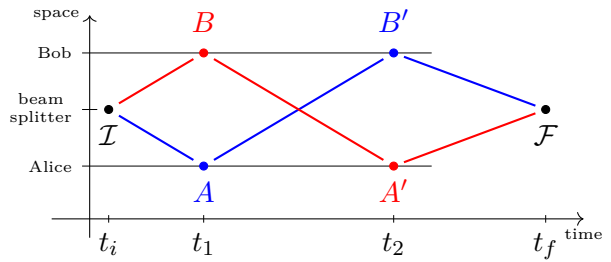
\begin{figure}[h!]
\begin{center}
\begin{tikzpicture}[x={(0cm,1cm)},y={(1cm,0cm)},scale=0.50]

\draw[->] (3.1,0) -- (9,0);
\node[anchor=east] at (9,0) {\tiny space};

\draw[->] (3.6,-1) -- (3.6,13);
\node[anchor=north] at (3.6,13) {\tiny time};
\draw[very thin] (3.5,0.5) -- (3.7,0.5);
\node[anchor=north] at (3.5,0.5) {$t_i$};
\draw[very thin] (3.5,3) -- (3.7,3);
\node[anchor=north] at (3.5,3) {$t_1$};
\draw[very thin] (3.5,8) -- (3.7,8);
\node[anchor=north] at (3.5,8) {$t_2$};
\draw[very thin] (3.5,12) -- (3.7,12);
\node[anchor=north] at (3.5,12) {$t_f$};

\draw[very thin] (5,-0.2) -- (5,9);
\node[anchor=east] at (5,-0.2) {\tiny Alice};

\draw[very thin] (8,-0.2) -- (8,9);
\node[anchor=east] at (8,-0.2) {\tiny Bob};

\node at (6.5,0.5) (splitterSi) {};
\node at (6.5,12) (splitterSf) {};
\draw[very thin] (6.5,-0.2) -- (6.5,0.2);
\node[anchor=east] at (6.5,-0.2) {$\substack{ \text{\tiny beam} \\ \text{\tiny splitter} }$};
\filldraw[black] (splitterSi) circle (3pt) node[anchor=north] {$\cI\strut$};
\filldraw[black] (splitterSf) circle (3pt) node[anchor=north] {$\cF\strut$};

\node at (5,3) (bluegateA) {};
\node at (8,8) (bluegateBprime) {};
\filldraw[blue] (bluegateA) circle (3pt) node[anchor=north] {$A\strut$};
\filldraw[blue] (bluegateBprime) circle (3pt) node[anchor=south] {$B^\prime\strut$};

\node at (8,3) (redgateB) {};
\node at (5,8) (redgateAprime) {};
\filldraw[red] (redgateB) circle (3pt) node[anchor=south] {$B\strut$};
\filldraw[red] (redgateAprime) circle (3pt) node[anchor=north] {$A^\prime\strut$};

\draw[thick,blue] (splitterSi) -- (bluegateA) -- (bluegateBprime) -- (splitterSf);

\draw[thick,red] (splitterSi) -- (redgateB) -- (redgateAprime) -- (splitterSf);


\end{tikzpicture}
\end{center}
\caption{\blue{Spacetime diagram, as well as the circuit representation, of the $4$-event implementation of the quantum switch.}}
\label{sl:pet}
\end{figure}

The realisations of the quantum switch are performed in table-top experiments in the gravitational field of the Earth, and can be for all practical purposes considered as being performed in flat Minkowski spacetime. In such experiments, Alice performs the unitary $U$ in her localised laboratory, and Bob performs $V$ in his separate localised laboratory, such that both are stationary with respect to each other and the Earth. The operations are applied on a single particle that arrives from the beam splitter, in a superposition of trajectories towards Alice and Bob, and, upon the exchange between the two agents, is finally recombined on the same beam splitter (for simplicity, we chose one beam splitter, but the whole analysis equally holds for two spatially separated beam splitters), and then measured. Below, we present a spacetime diagram of this experimental realisation of the quantum switch, which
also represents a circuit of the implementation scheme \blue{(see Figure~\ref{sl:pet})}.


Black horizontal lines represent world lines for Alice and Bob, as well as the global time coordinate line at the bottom. The black vertical line represents global space coordinate line. Quantum gates are represented by big dots. The composite gate $\mathcal{I}$ consists of the two preparation gates and the initial beam splitter gate, while $\mathcal{F}$ consists of the final beam splitter gate and the target gates that perform the final measurements (for details, see Appendix \ref{sec:AppFourEvent}). For simplicity, from now on we omit writing the labels of the nodes and keep only the labels of the corresponding circuit gates. The two histories of the particle exchanged between Alice and Bob, representing Alice's and Bob's wires, are full lines coloured in blue and red, respectively.

From the diagram we can see that in the blue history we have the following chain of gates
\begin{equation}
\label{eq:events_1}
\mathcal{I}\orderc A \orderc B^\prime \orderc \mathcal{F}\,,
\end{equation}
while for the red history we have
\begin{equation}
\label{eq:events_2}
\mathcal{I}\orderc B \orderc A^\prime \orderc \mathcal{F}\,.
\end{equation}
In total, there are four spacetime events involving Alice's and Bob's actions on the particle (gates), namely $A$, $B$, $A^\prime$ and $B^\prime$. \blue{Thus, we call the above diagram the ``$4$-event diagram''. This setup was already discussed in the literature (see the very end of the Supplementary Notes of~\cite{mac:rie:spe:res}).}

In order to compare the cases of the quantum and the gravitational switches, it would be interesting to analyse the two examples within recently introduced powerful {\em process matrix} formalism~\cite{ore:cos:bru:12}. To do so, one needs to formulate the formalism involving the vacuum state (see Appendix \ref{sec:AppI} for details). The straightforward application of the formalism to the $4$-event case is in full accord with the experimental results, as demonstrated in Appendix~\ref{sec:AppFourEvent}.

\subsection{3-event process}
\label{sec:3-event}

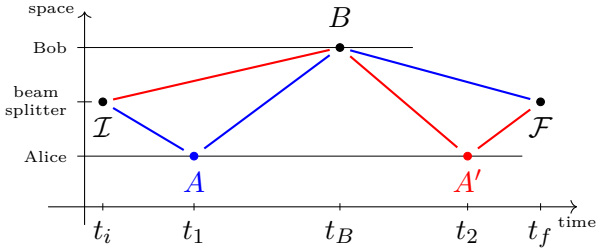
\begin{figure}[h!]
\begin{center}
\begin{tikzpicture}[x={(0cm,1cm)},y={(1cm,0cm)},scale=0.48]
\draw[->] (3.1,0) -- (9,0);
\node[anchor=east] at (9,0) {\tiny space};

\draw[->] (3.6,-1) -- (3.6,13.5);
\node[anchor=north] at (3.6,13.5) {\tiny time};
\draw[very thin] (3.5,0.5) -- (3.7,0.5);
\node[anchor=north] at (3.5,0.5) {$t_i$};
\draw[very thin] (3.5,3) -- (3.7,3);
\node[anchor=north] at (3.5,3) {$t_1$};
\draw[very thin] (3.5,7) -- (3.7,7);
\node[anchor=north] at (3.5,7) {$t_B$};
\draw[very thin] (3.5,10.5) -- (3.7,10.5);
\node[anchor=north] at (3.5,10.5) {$t_2$};
\draw[very thin] (3.5,12.5) -- (3.7,12.5);
\node[anchor=north] at (3.5,12.5) {$t_f$};

\draw[very thin] (5,-0.2) -- (5,12);
\node[anchor=east] at (5,-0.2) {\tiny Alice};

\draw[very thin] (8,-0.2) -- (8,9);
\node[anchor=east] at (8,-0.2) {\tiny Bob};

\node at (6.5,0.5) (splitterSi) {};
\node at (6.5,12.5) (splitterSf) {};
\draw[very thin] (6.5,-0.2) -- (6.5,0.2);
\node[anchor=east] at (6.5,-0.2) {$\substack{ \text{\tiny beam} \\ \text{\tiny splitter} }$};
\filldraw[black] (splitterSi) circle (3pt) node[anchor=north] {$\cI\strut$};
\filldraw[black] (splitterSf) circle (3pt) node[anchor=north] {$\cF\strut$};

\node at (5,3) (bluegateA) {};
\node at (8,7) (gateB) {};
\filldraw[blue] (bluegateA) circle (3pt) node[anchor=north] {$A\strut$};
\filldraw[black] (gateB) circle (3pt) node[anchor=south] {$B\strut$};

\node at (5,10.5) (redgateAprime) {};
\filldraw[red] (redgateAprime) circle (3pt) node[anchor=north] {$A^\prime\strut$};

\draw[thick,blue] (splitterSi) -- (bluegateA) -- (gateB) -- (splitterSf);

\draw[thick,red] (splitterSi) -- (gateB) -- (redgateAprime) -- (splitterSf);


\end{tikzpicture}
\end{center}
\caption{\blue{Spacetime diagram, as well as the circuit representation, of the $3$-event implementation of the quantum switch.}}
\label{sl:sest}
\end{figure}

\blue{One can imagine that instead of two, one of the agents implements only one gate.} For example, by conveniently choosing the velocity of the particle along its trajectory between Alice and Bob, we can identify Bob's two gates,
\begin{equation} \label{IdentifikacijaBgejtova}
B \equiv B^\prime\,.
\end{equation}
We thus arrive to the new spacetime diagram and the associated circuit, called the ``$3$-event diagram'' \blue{(see Figure~\ref{sl:sest})}.

Now, the obvious question is the following --- can we, in addition to (\ref{IdentifikacijaBgejtova}), impose also that
\begin{equation} \label{IdentifikacijaAgejtova}
A \equiv A^\prime\,,
\end{equation}
i.e., also identify Alice's gates into a single spacetime event? In flat Minkowski spacetime, the answer is negative. Namely, by simply looking at the $3$-event diagram one can see that the trajectory of the particle between Alice and Bob would require either superluminal speed, or backwards-in-time trajectory in at least one history (note that the diagram assumes that light propagates along the lines that form the $45^{\circ}$ angle with the coordinate axes). This is also seen directly from inequalities~\eqref{eq:events_1} and~\eqref{eq:events_2}: identifying both $A\equiv A^\prime$ and $B\equiv B^\prime$ would lead to requiring that {\em both} $A\orderc B$ and $B \orderc A$ are satisfied, i.e., $A \equiv B$. As guaranteed by our Theorem from Section~\ref{sec:causal_orders}, in a curved spacetime it is also impossible to make both identifications (\ref{IdentifikacijaBgejtova}) and (\ref{IdentifikacijaAgejtova}), at least if spacetime were globally hyperbolic. Finally, as in the $4$-event case, here also the process matrix formalism is consistent with the experimental results, see Appendix \ref{sec:AppThreeEvent}.

\subsection{\label{sec:gravitationalswitch} 2-event process --- gravitational switch}

Despite the conclusion of the previous subsection, within the framework of quantum gravity one is allowed to construct superpositions of different gravitational field configurations, leading to superpositions of different causal structures for the spacetime manifold. \red{The assumption of superpositions of different gravitational field configurations is common to all models of QG. Other than that, we will not have any additional assumptions, and thus our approach does not depend on any particular QG model.}

In what follows, \red{for the sake of concreteness,} we assume the ``traditional'' approach to the formulation of the QG formalism. Namely, we assume that there exists a smooth $4D$ manifold, called {\em spacetime}, and denoted as $\cM$. Quantum fields, including the gravitational field, live on top of $\cM$. The gravitational field is described either via the metric or via some other degrees of freedom (for example, tetrads and spin connection), such that the metric is a function of these. We call this kind of construction ``traditional'' because it represents a minimal deviation from the mathematical structure of quantum field theory (QFT) in flat Minkowski spacetime, in the sense of preserving the underlying manifold structure. A QG model implementing this approach is, for example, the asymptotic safety framework \cite{AsympSafetyReview}. Of course, we do not aim to provide a full-fledged model of QG, but rather to only specify the status of the manifold structure within it. As an alternative, in Subsection \ref{sec:relationalism}, we will discuss the relational framework of QG in which the manifold structure does not exist a priori, but is emergent from relational properties of quantum fields themselves. Finally, note that the discussion of the flat-spacetime cases in the previous sections implicitly assumes the traditional point of view on spacetime manifold. Nevertheless, it has to be compatible with the semiclassical limit of any viable QG model.

As a consequence of the superposition of causal structures in QG, it is possible to achieve a {\em gravitational switch}, which implements the same quantum switch as described above, with a circuit consisting (in addition to the initial and final gates $\mathcal{I}$ and $\mathcal{F}$) of only two gates: the Alice's gate $A$ that applies $U$, and Bob's gate $B$ that applies $V$. Superposing two gravity-matter states, such that in the first the spacetime geometry (described by the metric tensor $g_{0}$) establishes the causal structure
\begin{equation} \label{eq:PlaviRedosledDogadjajaUprostorvremenu}
\mathcal{I} \orderm^{g_0} A \orderm^{g_0} B \orderm^{g_0} \mathcal{F}\,,
\end{equation}
while in the second (described by the metric $g_1$) it is
\begin{equation} \label{eq:CrveniRedosledDogadjajaUprostorvremenu}
\mathcal{I} \orderm^{g_1} B \orderm^{g_1} A \orderm^{g_1} \mathcal{F}\,,
\end{equation}
the overall circuit applies operations $U_0 = UV$ and $U_1 = VU$, conditioned on the state of gravity. \red{As a side note, it is clear from (\ref{eq:PlaviRedosledDogadjajaUprostorvremenu}) and (\ref{eq:CrveniRedosledDogadjajaUprostorvremenu}) that superpositions of the spacetime causal orders can occur only in the framework of quantum gravity.}

Such a switch was previously introduced by Zych {\em et al.}~\cite{zyc:cos:pik:bru:17}, in the context of two spacetimes which are solutions of the Einstein equations. In their proposal, the beam splitter acted {\em only} on the gravitational degree of freedom (and the accompanied source, the planet), while leaving the rest of the matter, in particular the particle, Alice and Bob, unaffected. Upon the final beam splitter recombination, the matter is left in an {\em incoherent} mixture of two states proportional to $\akomut{U}{V}\ket{\Psi}$ and $\komut{U}{V} \ket{\Psi}$. \blue{Subsequently, the mass (along with its gravitational degrees of freedom) is being measured in the superposition basis. Upon a post-selection conditioned on the outcome of the measurement, the matter is again in a pure state.}

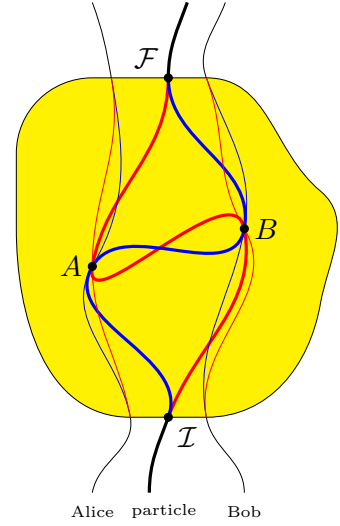
\begin{figure}[!ht]
\begin{center}
\begin{tikzpicture}[scale=0.5]

\draw[very thin, fill=yellow] (5,2) to [out=0, in=180 ] (7,2) to [out=0, in=260] (10,5) to [out=80, in=320] (10,8) to [out=140, in=0] (7,11) to [out=180, in=0] (4,11) to [out=180, in=90] (2,9) to [out=270, in=90] (2,7) to [out=270, in=180] (5,2);

\node at (5.5,-0.5) {$\quad$\tiny particle};
\draw[very thick, black] (5.5,0) to [out=90 , in=250] (6,2);
\draw[very thick, black] (6,11) to [out=90 , in=250] (6.5,13);

\draw[very thick, red] (6,2) to [out=70, in=280] (8,7) to [out=100 , in=260] (4,6) to [out=80 , in=270] (6,11); 

\draw[very thick, blue] (6,2) to [out=70, in=240] (4,6) to [out=60 , in=260] (8,7) to [out=80 , in=270] (6,11); 

\node at (4,-0.5) {\tiny Alice};
\draw[very thin, black] (4,0) to [out=80, in=280] (5,2);
\draw[very thin, black] (4.5,11) to [out=100, in=290] (4,13);

\draw[very thin, blue] (5,2) to [out=100, in=240] (4,6) to [out=60, in=280] (4.5,11);

\draw[very thin, red] (5,2) to [out=100, in=270] (4,6) to [out=90, in=280] (4.5,11);

\node at (8,-0.5) {\tiny Bob};
\draw[very thin, black] (8,0) to [out=90, in=280] (7,2);
\draw[very thin, black] (7,11) to [out=110, in=250] (7.5,13);

\draw[very thin, blue] (7,2) to [out=100, in=260] (8,7) to [out=80, in=290] (7,11);

\draw[very thin, red] (7,2) to [out=100, in=300] (8,7) to [out=120, in=290] (7,11);

\filldraw[black] (4,6) circle (3pt) node[anchor=east] {$A$};
\filldraw[black] (8,7) circle (3pt) node[anchor=west] {$B$};
\filldraw[black] (6,2) circle (3pt) node[anchor=north west] {$\mathcal{I}$};
\filldraw[black] (6,11) circle (3pt) node[anchor=south east] {$\mathcal{F}$};

\end{tikzpicture}
\end{center}
\caption{\blue{Spacetime diagram of the $2$-event implementation of the gravitational switch. Note that formally this is not a circuit diagram, as the control wire, implemented by the state of the gravitational field in the yellow region, is missing.}}
\label{sl:sedam}
\end{figure}

Another way to obtain a genuine superposition of two different causal orders is by using a spatially delocalised beam splitter, that acts on both gravitational and matter fields. This can be depicted by the following $2$-event diagram \blue{(see Figure~\ref{sl:sedam})}.

The yellow region in this diagram represents a compact piece of spacetime where the gravitational field is in a superposition of the two distinct states, and plays the role of the control degree of freedom. Along the boundary of that region, both gravitational configurations smoothly join into a single configuration outside. The boundary of the yellow region thus acts as a beam splitter for anything that enters, and again (in the recombining role) for anything that exits. Therefore, all worldlines (namely, of Alice, Bob and the particle) are doubled inside the yellow region. The blue and red colours represent their spacetime trajectories in two different gravitational field backgrounds, respectively.

We model our gravitational switch such that the overall output state is the product between the state of the gravitational field and the state of the particle. The state of the particle is of the form $(\alpha UV + \beta VU)\ket{\Psi}$\blue{, obtained without performing final selective measurement.} In particular, in order to compare it with the other quantum switch realizations, we choose either $\akomut{U}{V}\ket{\Psi}$ or $\komut{U}{V}\ket{\Psi}$. In order to achieve this, the gravitational switch should act upon {\em all} degrees of freedom, both gravitational and matter. \blue{Note that our gravitational switch does require certain fine tuning, in the sense that the whole, delocalised beam splitter, that acts nontrivially on the whole joint gravity-matter system, is designed for the particular pair of operations applied by Alice and Bob: only for those operations, the beam splitter will output the product state between gravity and matter. Otherwise, the output will be the entangled gravity-matter state, like in the cases of the optical quantum switch and the gravitational switch introduced by Zych et al. (before the final selective measurement). Still, the process matrix describing the gravitational switch itself is independent of the choice of the gate operations of the agents.} See Appendix \ref{sec:AppTwoEvent} for details.

The question whether this kind of diagram is admissible in some theory of quantum gravity is nontrivial, and model dependent, on several grounds. First, it is impossible to construct this diagram by superposing two classical configurations of gravitational field, such that each configuration satisfies Einstein equations. The reason is simple --- assuming that the gravitational field is specified outside the yellow region, Einstein equations have a unique solution (up to diffeomorphism symmetry) for the compact yellow region, given such a boundary condition. Therefore, one cannot have two different solutions to superpose inside. The only two options are to either superpose one on-shell and one off-shell configuration of gravity, or two off-shell configurations. This scenario can arguably be considered within the path integral framework for quantum gravity.

Second, the question of the particle trajectory is nontrivial. Namely, given one gravitational configuration in which the particle has the spacetime causal structure (\ref{eq:PlaviRedosledDogadjajaUprostorvremenu}), corresponding to the blue history, it is not obvious that there can exist another gravitational configuration (with the same boundary conditions at the edge of the yellow region), in which the particle has the spacetime causal structure (\ref{eq:CrveniRedosledDogadjajaUprostorvremenu}), corresponding to the red history. Even if one admits arbitrary off-shell configurations of gravity, it may turn out that the order of events inside the yellow region must be fixed by the boundary conditions. The only viable way to answer this question is to try and construct an explicit example of two geometries implementing (\ref{eq:PlaviRedosledDogadjajaUprostorvremenu}) and (\ref{eq:CrveniRedosledDogadjajaUprostorvremenu}) for the same boundary conditions. Numerical investigations are underway to explore this possibility.

\section{\label{sec:distinguishing}Distinguishing 2-, 3-, and 4-event realisations of the quantum switch}

\blue{In a number of both theoretical proposals~\cite{chi:dar:per:val:13,ore:cos:bru:12,ara:bra:cos:fei:gia:bru:15,bad:ara:bru:que:19,gue:rub:bru:18}, as well as experimental realisations~\cite{pro:etal:15,rub:roz:fei:ara:zeu:pro:bru:wal:17,rub:roz:mas:ara:zyc:bru:wal:17} of the quantum switch, it is claimed that they feature genuine superpositions of causal orders. The reason for this is the introduction of an alternative, operational notion of the event, which differs from a spacetime point. The motivation for this lies in the claim that the individual spacetime points $A$ and $A^\prime$ (and $B$ and $B^\prime$) do not have an operational meaning. In words of the authors of \cite{pro:etal:15} (see the Discussion section):
\begin{quote}
{\em ``The results of the experiment confirm that such [which way] information is not available anywhere and that the interpretation of the experiment in terms of four, causally-ordered events cannot be given any operational meaning. If, on the other hand, one requires events to be defined operationally, in terms of measurable interactions with physical systems [...], then the experiment should be described in terms of only two events --- a single use of each of the two gates.''}
\end{quote}
While it is obvious that the mentioned which-way information is not available in the quantum switch experiment, in what follows we argue that this does not imply that one cannot give an operational meaning to spacetime points, even in the context of the quantum switch in classical geometries.}

Below, we first present a critical analysis of the \blue{arguments behind introducing the operational notion of event}. Then, we show how one can experimentally, at least in principle, distinguish 2-, 3-, and 4-event realisations of the quantum switch.

It is the operational approach to understanding spacetime, applied within the framework of relationalism (see Section~\ref{sec:discussion} for a detailed discussion of the relation between the two frameworks), that is arguably the main argument \blue{for introducing the alternative notion of an event. This new notion of an event gives rise to the superposition of respective causal orders in the realisations of the quantum switch even in classical spacetimes.} Assuming that the smooth (classical) spacetime is an emergent phenomenon, in the operational approach one considers ``closed laboratories'' \cite{ore:cos:bru:12} as the primal entities within which one can locally apply standard quantum mechanics, while their connections form the relations from which the spacetime emerges. Indeed, it seems that the process matrix formalism was developed precisely with this idea in mind: to be a mathematical tool in analysing the emergence of the spacetime through the relations between the closed laboratories. We would like to note that, as shown in Appendices~\ref{sec:AppFourEvent},~\ref{sec:AppThreeEvent} and~\ref{sec:AppTwoEvent}, the mentioned formalism is \blue{also} fully applicable within the standard formulation of quantum mechanics in classical Minkowski spacetime.

Given that in the case of coherent superpositions of the two paths (a particle first goes to Alice, then to Bob, and vice versa) it is not possible to know which of the two has actually been taken, one may conclude that one cannot distinguish between \blue{spacetime} events $A$ and $A^\prime$, and that the two are operationally given by the single action of a {\em spatially} localised laboratory. However, this point of view is at odds with our understanding of the ordinary double slit experiment. Namely, by exchanging the roles of time and space, and following the above logic, applied to the case of the standard double slit experiment, one could analogously conclude that, since in order to obtain the interference pattern at the screen one must not (and thus cannot) learn which slit the particle went through, the two slits represent one and the same \blue{operational} ``lab'', and one \blue{operational} point (region) in space.

\blue{Let us explain our argument in slightly more detail. Consider first the optical quantum switch. Here, a particle passes through Alice's lab, described by the two spacetime points, $(x_A, t)$ and $(x_A, t')$. Any attempt to distinguish the times $t$ and $t'$ at which the particle passes through Alice's lab would destroy the superposition. Consider now the standard double slit experiment. Here, a particle passes through the two slits, described by the two spacetime points, $(x_L,t)$ and $(x_R,t)$. Any attempt to distinguish the positions of the slits $x_L$ and $x_R$ through which the particle passes would destroy the superposition. Note that by exchanging the roles of space and time, the descriptions of the above two situations are essentially identical.

According to the operational approach, as a consequence of the above, one should describe Alice's actions in the optical quantum switch in terms of only one operational event. Thus, analogously, one should also describe the particle passing through the slits in terms of only one operational event. However, such interpretation of the double slit experiment is, to the best of our knowledge, absent from the literature.}

\blue{Note also that} the 3-event realisation of the quantum switch offers a natural alternative interpretation of this phenomenon, as a well known {\em time double slit experiment}~\cite{bau:14}. Indeed, the two events (gates) $A$ and $A^\prime$ play the role of the two time-like slits, while the event (gate) B separates the two in the same way the closed shutter separates the two time-like slits in the time double slit experiment. This comes as no surprise: quantum superpositions are in general accompanied by the interference effects, and the quantum switch is, as already emphasised in Section~\ref{sec:quantum_switch}, just another instance of a superposition of two different states of the standard quantum mechanics \blue{in Minkowski spacetime}.

\blue{The operational interpretation of identifying the events $A$ and $A^\prime$ in the current experimental realisations of the quantum switch indeed seems to be a tempting proposal. Nevertheless, we would like to point out that in fact it does not resolve any open problem. In addition, being similar to Mach's ideas, it too may be at odds with the theory of general relativity (GR), see Subsection~\ref{sec:mach} for a detailed discussion.}

\subsection{Distinguishing by decohering the particle}
\label{sec:distinguishing_decohering}

\blue{In the above quote from~\cite{pro:etal:15}, the authors claim that in order to directly distinguish points $A$ and $A^\prime$ (as well as $B$ and $B^\prime$), one {\em must} destroy the superposition in the apparatus. Conversely, being unable to distinguish those points in any experiment that maintains superposition and realises the quantum switch, one cannot give them operational meaning. Therefore, those spacetime points are redundant in the theory, and each pair should be replaced by a single operational event. In this subsection, we discuss this type of an argument. In the next, we give an explicit example of an observable that does distinguish such spacetime points without obtaining the which way information.}

\blue{Let us study one concrete way of distinguishing the mentioned pairs of points, which decoheres the particle.} For simplicity, we will analyse the 4- and 2-event cases only. To this end, we will introduce a third agent, Alice's and Bob's {\em Friend}. At each run of the quantum switch experiment, Alice will, independently and at random, decide whether just to apply her operation onto the particle, or in addition to that, send a photon to Friend. The same holds for Bob. In 25\% of the cases, both agents just perform their respective operations, thus performing the quantum switch. Next, in the 25\% of the cases, both agents decide, in addition to applying their respective operations, to send the photons to Friend, who detects them in his spatially localised lab. The remaining 50\% of the cases are essentially the same as the previous ones, so for simplicity we omit their analysis.

\begin{figure}[h!]
\begin{center}
\begin{tikzpicture}[x={(0cm,1cm)},y={(1cm,0cm)},scale=0.47]
\draw[->] (3,0) -- (10,0);
\node[anchor=east] at (10,0) {\tiny space};

\draw[->] (3.6,-1) -- (3.6,14);
\node[anchor=north] at (3.6,14) {\tiny time};
\draw[very thin] (3.5,0.5) -- (3.7,0.5);
\node[anchor=north] at (3.5,0.5) {$t_i$};
\draw[very thin] (3.5,3) -- (3.7,3);
\node[anchor=north] at (3.5,3) {$t_1$};
\draw[very thin] (3.5,8) -- (3.7,8);
\node[anchor=north] at (3.5,8) {$t_2$};
\draw[very thin] (3.5,13) -- (3.7,13);
\node[anchor=north] at (3.5,13) {$t_f$};

\draw[very thin] (9,-0.2) -- (9,13);
\node[anchor=east] at (9,-0.2) {\tiny Friend};

\draw[very thin] (5,-0.2) -- (5,9);
\node[anchor=east] at (5,-0.2) {\tiny Alice};

\draw[very thin] (8,-0.2) -- (8,9);
\node[anchor=east] at (8,-0.2) {\tiny Bob};

\node at (6.5,0.5) (splitterSi) {};
\node at (6.5,13) (splitterSf) {};
\filldraw[black] (splitterSi) circle (3pt) node[anchor=north] {$\mathcal{I}\strut$};
\filldraw[black] (splitterSf) circle (3pt) node[anchor=north] {$\mathcal{F}\strut$};

\node at (5,3) (bluegateA) {};
\node at (8,8) (bluegateBprime) {};
\node at (9,7) (bluegateFA) {};
\node at (9,9) (bluegateFBprime) {};
\filldraw[blue] (bluegateA) circle (3pt) node[anchor=north] {$A\strut$};
\filldraw[blue] (bluegateBprime) circle (3pt) node[anchor=north] {$B^\prime\strut$};
\filldraw[blue] (bluegateFA) circle (3pt) node[anchor=south] {$F_{A}\strut$};
\filldraw[blue] (bluegateFBprime) circle (3pt) node[anchor=south] {$F_{B^\prime}\strut$};

\draw[thick,blue] (splitterSi) -- (bluegateA) -- (bluegateBprime) -- (splitterSf);

\draw[dotted,blue] (bluegateA) -- (bluegateFA);
\draw[dashed,blue] (bluegateBprime) -- (bluegateFBprime);

\node at (8,3) (redgateB) {};
\node at (5,8) (redgateAprime) {};
\node at (9,4) (redgateFB) {};
\node at (9,12) (redgateFAprime) {};
\filldraw[red] (redgateB) circle (3pt) node[anchor=north] {$B\strut$};
\filldraw[red] (redgateAprime) circle (3pt) node[anchor=north] {$A^\prime\strut$};
\filldraw[red] (redgateFB) circle (3pt) node[anchor=south] {$F_{B}\strut$};
\filldraw[red] (redgateFAprime) circle (3pt) node[anchor=south] {$F_{A^\prime}\strut$};

\draw[thick,red] (splitterSi) -- (redgateB) -- (redgateAprime) -- (splitterSf);

\draw[dotted,red] (redgateAprime) -- (redgateFAprime);
\draw[dashed,red] (redgateB) -- (redgateFB);

\end{tikzpicture}
\end{center}
\caption{\blue{Distinguishing spacetime points by decohering the particle in the $4$-event quantum switch. The dotted (dashed) lines represent photons sent by Alice (Bob) to Friend.}}
\label{sl:osam}
\end{figure}
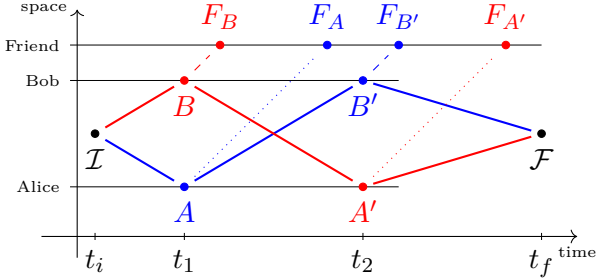

First, we present the spacetime diagram of the 4-event quantum switch for the case when the agents decide to send the photons to Friend \blue{(see Figure~\ref{sl:osam})}.

The photons coming from Alice are dotted, while the photons coming from Bob are dashed. By knowing the geometry of the whole experiment, Friend would  be able to measure, in a generic setup, four {\em different} times of the photon arrivals: $t_A$ and $t_{A^\prime}$ for spacetime points $F_A$ and $F_{A^\prime}$, and two more for the photons sent by Bob.

\begin{figure}[!ht]
\begin{center}
\begin{tikzpicture}[scale=0.5]

\draw[very thin, fill=yellow] (5,2) to [out=0, in=180 ] (7,2) to [out=0, in=260] (10,5) to [out=80, in=320] (10,8) to [out=140, in=0] (7,11) to [out=180, in=0] (4,11) to [out=180, in=90] (2,9) to [out=270, in=90] (2,7) to [out=270, in=180] (5,2);

\draw[very thin] (0,0) to [out=60, in=280 ] (0,8) to [out=100, in=260] (0,11) to [out=80, in=280] (0,13);
\node at (0,-0.5) {\tiny Friend};

\draw[dashed, blue] (8,7) to [out=160, in=330] (2,9);
\draw[dashed, red] (8,7) to [out=100, in=330] (2,9);
\draw[dashed, black] (2,9) to [out=150, in=350] (0,11);

\draw[dotted, blue] (4,6) to [out=180, in=350] (2,7);
\draw[dotted, red] (4,6) to [out=100, in=350] (2,7);
\draw[dotted, black] (2,7) to [out=170, in=300] (0,8);

\node at (5.5,-0.5) {$\quad$\tiny particle};
\draw[very thick, black] (5.5,0) to [out=90 , in=250] (6,2);
\draw[very thick, black] (6,11) to [out=90 , in=250] (6.5,13);

\draw[very thick, red] (6,2) to [out=70, in=280] (8,7) to [out=100 , in=260] (4,6) to [out=80 , in=270] (6,11); 

\draw[very thick, blue] (6,2) to [out=70, in=240] (4,6) to [out=60 , in=260] (8,7) to [out=80 , in=270] (6,11); 

\node at (4,-0.5) {\tiny Alice};
\draw[very thin, black] (4,0) to [out=80, in=280] (5,2);
\draw[very thin, black] (4.5,11) to [out=100, in=290] (4,13);

\draw[very thin, blue] (5,2) to [out=100, in=240] (4,6) to [out=60, in=280] (4.5,11);

\draw[very thin, red] (5,2) to [out=100, in=270] (4,6) to [out=90, in=280] (4.5,11);

\node at (8,-0.5) {\tiny Bob};
\draw[very thin, black] (8,0) to [out=90, in=280] (7,2);
\draw[very thin, black] (7,11) to [out=110, in=250] (7.5,13);

\draw[very thin, blue] (7,2) to [out=100, in=260] (8,7) to [out=80, in=290] (7,11);

\draw[very thin, red] (7,2) to [out=100, in=300] (8,7) to [out=120, in=290] (7,11);

\filldraw[black] (4,6) circle (3pt) node[anchor=east] {$\tilde A$};
\filldraw[black] (8,7) circle (3pt) node[anchor=west] {$\tilde B$};
\filldraw[black] (0,8) circle (3pt) node[anchor=east] {$\tilde{F}_A$};
\filldraw[black] (0,11) circle (3pt) node[anchor=east] {$\tilde{F}_B$};
\filldraw[black] (6,2) circle (3pt) node[anchor=north west] {$\mathcal{I}$};
\filldraw[black] (6,11) circle (3pt) node[anchor=south east] {$\mathcal{F}$};

\end{tikzpicture}
\end{center}
\caption{\blue{Distinguishing spacetime points by decohering the particle in the $2$-event gravitational switch. The dotted (dashed) lines represent photons sent by Alice (Bob) to Friend.}}
\label{sl:devet}
\end{figure}
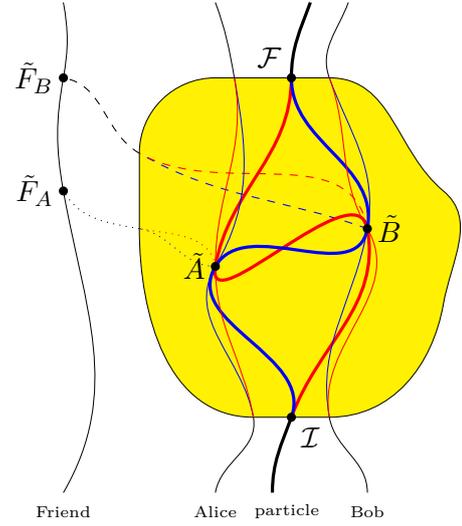

On the other hand, in the case of the 2-event gravitational switch realisation, Friend would detect only two times of the photons' arrival. Below, we extend the diagram of the gravitational switch we introduced in Section~\ref{sec:gravitationalswitch}, by adding the photons sent to Friend. In order to indicate the fact that the events $A$ and $A^\prime$, etc., are in this setup indeed identified, we write the tilde over the corresponding letters $A$, $B$ and $F$ \blue{(see Figure~\ref{sl:devet})}.

Clearly, the two situations are experimentally distinguishable.

\blue{Nevertheless, as noted in \cite{pro:etal:15}, one might argue that, since the photons sent to Friend in the $4$-event case decohere the particle in the switch, this situation does not correspond to the experiment in which the coherence is maintained. Therefore, in the latter, the pair of spacetime events $A$ and $A^\prime$ still ought to be substituted with a single operational event (and analogously for $B$ and $B^\prime$).}

\blue{However, even if instead of spacetime points one decides to talk about operational events, such a framework should still be consistent with the experimentally tested theories, GR in particular. According to GR, in flat spacetime (or in any classical configuration of the gravitational field), regardless of whether we decohere the particle or not, both experiments feature four spacetime points, such that $A$ and $B$ (as well as $A^\prime$ and $B^\prime$) can be considered to be simultaneous (see Figure~\ref{sl:osam}). Therefore, the time of execution of both experiments is $\delta t = t_2 - t_1 + C$, where $C\equiv (t_1 - t_i) + (t_f - t_2)$.  Note that the time period $t_2-t_1$ represents the travel time of the particle from one laboratory to the other, and is therefore strictly positive.

From the operational point of view, the decohered version of the experiment also features four operational events, and is thus manifestly consistent with the GR description. Note that a decohered version of the switch still features only two events {\em per run}: in a classical mixture between ``Alice's event before Bob's event'' and ``Bob's event before Alice's event'' each run features just two events, and the duration of the overall experiment in each run is the time between the two events of that run (plus the above constant $C$).

On the other hand, if the coherence were maintained, the operational point of view features only two operational events, one per laboratory. Then, the total time of execution of the experiment ought to be $\delta \tau = 0 + C$, which is clearly different from the GR prediction. The total time of execution of the quantum switch experiment is a measurable quantity. This means that one can easily determine whether this time is $\delta \tau$ or $\delta t$. The former outcome invalidates GR, which would necessitate the formulation of an alternative theory. Note that in this case, a sheer decision to either decohere a particle or not would allow agents to influence the time flow in their labs. Moreover, it raises the question of the time flow in nearby labs {\em isolated} from the experiment during its execution. The latter outcome poses the problem of the precise formulation of an {\em operational theory} such that the experiment which features only two operational events lasts precisely the same time as the experiment which features four operational events.
}

\subsection{Distinguishing without decohering the particle}
\label{sec:distinguishing_without_decohering}

In addition to the above argument, supported by the experimental setup presented in the previous subsection, by erasing the which way information it {\em is} possible for Friend to distinguish the 4-event and the 2-event realisations even when the ``full'' quantum switch is executed. For that, one needs to supply Friend with a photon non-demolition measurement. This is in principle possible to construct, although in practice a bit challenging. It thus might be technically easier to use some particles other than photons for sending signals to Friend.  

By agreeing in advance of the particular experimental setup, Friend would be able to predict the {\em distinct} times of arrival of the photons, $t_{F_A}$, $t_{F_{A^\prime}}$, $t_{F_B}$ and $t_{F_{B^\prime}}$ in the 4-event case, and $t_{\tilde{F}_A}$, $t_{\tilde{F}_B}$ in the 2-event case, thus defining the states of the two photons that arrive to his lab: $\ket{F_A, F_{B^\prime}}$ , $\ket{F_{A^\prime}, F_{B}}$, and $\ket{\tilde{F}_A, \tilde{F}_B}$, respectively. Let us define $\mathcal{H}_{A \prec B^\prime} = \sspan \{ \ket{F_A, F_{B^\prime}}\} $, $\mathcal{H}_{B \prec A^\prime} = \sspan \{ \ket{F_{A^\prime}, F_{B}}\}$, and $\mathcal{H}_{A \prec B \land B \prec A} = \sspan \{ \ket{\tilde{F}_A, \tilde{F}_B}\} $. Then, the relevant Hilbert space of the two photons is
\begin{equation}
\label{eq:friend_space}
\begin{array}{rl}
\mathcal{H}_{ph} = \mathcal{H}_{A \prec B^\prime} \oplus \mathcal{H}_{B \prec A^\prime} \oplus \mathcal{H}_{A \prec B \land B \prec A} \,.
\end{array}
\end{equation}
Let us define $P_<$, $P_>$ and $P_=$ as orthogonal projectors onto $\mathcal{H}_{A \prec B^\prime}$, $\mathcal{H}_{B \prec A^\prime}$ and $\mathcal{H}_{A \prec B \land B \prec A}$, respectively. One can then define a dichotomic photon non-demolition orthogonal observable performed by Friend on the two photons in his laboratory:
\begin{equation}
	\label{eq:friend_observable_1}
	M = 1 \cdot (P_< + P_>) + 0 \cdot P_= \, .
\end{equation}

Provided that the experimental setup is {\em either} that of the 4-event, or the 2-event type, such measurement would not change the state of the experimental setup (the interferometer, the particle in it, and the photons in the Friend's apparatus), while still leaking the information to Friend (via the measurement outcome) about the type of the quantum switch realisation. Finally, by performing the {\em quantum erasing} procedure~\cite{scu:dru:82, scu:eng:her:91}, the which way information is lost, and the final state of the particle is restored to a coherent superposition.

Let us examine this more formally. Let the two states of the particle in the quantum switch be $\ket{R}$ and $\ket{B}$, corresponding to the red and the blue trajectory, respectively. After $\mathcal{I}$, the state of the particle in the quantum switch is $\frac{1}{\sqrt{2}}(\ket{R} + \ket{B})$. As the particle passes through Alice's and Bob's labs, two photons are emitted, which arrive at the Friend's lab. The overall state of the particle and the two photons in the 2-event quantum switch is then
\begin{equation}
	\frac{1}{\sqrt{2}}\Big(\ket{R} + \ket{B}\Big)\ket{\tilde{F}_A, \tilde{F}_B}.
\end{equation}
The particle in the quantum switch is in superposition of the two paths, and it stays so upon measuring $M$ and obtaining the result 0.

On the other hand, the overall state of the particle and the two photons in the 4-event quantum switch is, upon the photons' arrival in the Friend's lab, given by
\begin{equation}
\frac{1}{\sqrt{2}} \Big(\ket{R}\ket{F_{A^\prime}, F_{B}} + \ket{B}\ket{F_{A}, F_{B^\prime}}\Big) \hphantom{aaaaaa}
\end{equation}
$$
\hphantom{aa} = \frac{1}{2\sqrt{2}}\Big[\Big(\ket{R} + \ket{B}\Big)\Big(\ket{F_{A^\prime}, F_{B}} + \ket{F_{A}, F_{B^\prime}}\Big)
$$
$$
\hphantom{aaaa} + \Big(\ket{R} - \ket{B}\Big)\Big(\ket{F_{A^\prime}, F_{B}} - \ket{F_{A}, F_{B^\prime}}\Big)\Big].
$$
     
The particle is now decohered by the two photons, and it remains so upon measuring $M$ and obtaining 1 as the result. Therefore, to erase the which way information, Friend has to perform an additional measurement in the basis
\begin{equation}
\label{eq:friend_observable_2}
	\ket\pm = \frac{1}{\sqrt{2}}\Big(\ket{F_{A^\prime}, F_{B}} \pm \ket{F_{A}, F_{B^\prime}}\Big),
\end{equation}
thus collapsing the state of the particle in one of the two pure states
\begin{equation}
	\frac{1}{\sqrt{2}}\Big(\ket{R} \pm \ket {B}\Big).\end{equation}
Knowing the outcome of the measurement of $M$, Friend can post-select the output of the particle coming out of the quantum switch. Alternatively, in the case of obtaining the $\ket -$ result, Friend can change the relative phase between the two of the particle's superposed states.

\subsection{\label{sec:otherswitches}Other types of gravitational switches}

It is important to note that the framework of QG also allows for the construction of $3$- and $4$-event switches, in addition to the $2$-event one. This is straightforward to see, for example by immersing the above $3$- or $4$-event spacetime diagram into a superposition of different geometries.

Moreover, all of these gravitational switches may give different outcomes when measuring the observable $M$, given by (\ref{eq:friend_observable_1}), followed by the quantum erasing procedure~\eqref{eq:friend_observable_2}. The criteria to necessarily obtain the outcome~$0$ are: (i) that the photons in red and blue histories meet at the boundary of the yellow region, and (ii) from that point on they recombine into a single photon history. Depending on the details of their construction, all gravitational switches either may or may not satisfy the criteria (i) and (ii). On the other hand, no quantum switch realisations in classical spacetimes with definite causal order could ever yield result $0$. Finally, we note that even though some of $2$-event gravitational switches may give the outcome $1$ when measuring $M$, it does not necessarily mean that there exist no other observable that could distinguish them from the $4$-event quantum switches in a classical geometry. This is a matter for further research.

Detailed graphical visualisations of various gravitational switches are presented in the Appendix \ref{app:gravitacioni_svicevi}.

\section{\label{sec:discussion}Relational approach to physics}

\blue{In the light of the operational framework, which suggests the substitution of the spacetime events $A$ and $A^\prime$ with a single operational event (and analogously for $B$ and $B^\prime$),} it is important to comment on one different but related approach to understanding spacetime, called relationalism. \blue{Note that by this promotion of operational events as fundamental entities that ought to replace and play the role of the spacetime events, effectively means the identification of $A$ with $A^\prime$, and $B$ with $B^\prime$.} In this section, we first present a historical review of the relational approach to physics. Then, we discuss the operational framework within the context of the modern approach to relationalism.

\subsection{Mach principle and the history of relationalism}
\label{sec:mach}

The idea of relationalism is an old one, it traces back at least as far as Decartes, and is very important in human thought, in particular in the history of physics. It was brought back to science by Mach in the second half of the XIX century (for an overview and history of the Mach principle and the relational approach to space, from its origins in ancient Greece, see for example~\cite{lic:mas:04} and the references therein). Based on the Leibniz ideas of a relational world, Mach formulated his famous Mach principle, an intuitively reasonable approach in analysing physics, and space(time) relations in particular. One of the main characteristics of the Mach principle is that (see~\cite{bla:02}, page 17):
\begin{quote}
{\em ``Space as such plays no role in physics; it is merely an abstraction from the totality of spatial relations between material objects.''}
\end{quote}
The same formulation can be found in~\cite{bon:sam:97}, slightly re-phrased as {\em ``Mach7: If you take away all matter, there is no more space.''} It is interesting to note that the authors attribute this formulation to A. S. Eddington~\cite{edd:21}, page 164.

As discussed at the beginning of Section~\ref{sec:distinguishing}, in the operational approach one attributes the ultimate existence to the ``closed laboratories'' only, while their mutual relations, epitomised by the process matrix, are then giving rise to higher level emergent entities. This clearly shows striking similarities between the Mach's and the operational approaches to space(time). 

Mach's ideas were crucial for Einstein in formulating the theory of relativity. And while many of Mach's predictions were indeed realised in the new theory, some of them were not. Mach's idea that the matter is the basic entity, and that by abstracting the relations between the objects the space emerges, led him to the following statement: if the matter in the universe were finite and had 3D rotational symmetry, it would be impossible to determine its angular momentum (indeed, even talking about it would have no meaning). This is a plausible idea. Nevertheless, it does not hold in general relativity (GR), where one can find two solutions of the Einstein equations for the isolated black hole (the stationary Schwarzschild solution and the rotating Kerr solution~\cite{mis:tho:the:73}). Moreover, while according to the Mach principle the matter completely determines the space, this is not the case in GR: not only that there exists a solution for the gravitational field in the absence of matter (when the stress-energy tensor $T$ is identically zero), but the solution is not unique, as it depends on the boundary conditions as well (i.e., flat Minkowski spacetime is not the only solution --- gravitational waves being a possible alternative~\cite{mis:tho:the:73}). This also holds for the general $T \neq 0$ case, as there too boundary conditions play an important role. Thus, matter does not fully determine the inertia, as should according to Mach principle, which states that the inertia of a massive body is given solely in terms of its relations with the other massive bodies.

Motivated by giving the ultimate reality to material objects only (closed laboratories in the case of the operational approach), Mach formulated the above list of claims. Nevertheless, they were later shown not to hold in GR. \blue{Provided the similarities between the Mach ideas and the operational approach, the latter might face similar problems as well. We thus believe that introducing the operationalist notion of an event should be accompanied by more elaborate proposals of new physical hypotheses and theories.} We hope that our discussion may serve as a small step towards achieving this goal.

\subsection{Modern approach to relationalism}
\label{sec:relationalism}

In contrast to the historical approach to relationalism and Mach's ideas, that sounded plausible at the time but ultimately failed with the development of GR, the more elaborate modern approach to relationalism is epitomised in the words of Carlo Rovelli (see Section 2.3 of~\cite{rov:04}):
\begin{quote}
{\em ``The world is made up of fields. Physically, these do not live on spacetime. They live, so to say, on one another. No more fields on spacetime, just fields on fields.''}
\end{quote}
In particular, the modern relational approach to spacetime defines a particular spacetime point by the physical processes that are ``happening at that point''. More technically, given an ordered set of classical fields $\phi \equiv (\phi_1,\dots,\phi_n )$ used to describe physics in a given classical theoretical framework, one traditionally starts from some spacetime point $\tilde{x}$ and evaluates the fields at that point, $\tilde{\phi}_i = \phi_i(\tilde x)$, obtaining an $n$-tuple of numbers $(\tilde\phi_1,\dots,\tilde\phi_n)$. The idea of relationalism does the opposite --- one starts from $n$-tuples of field values, and then defines a spacetime point using an $n$-tuple, $\tilde x \equiv (\tilde\phi_1,\dots,\tilde\phi_n)$, so that the same equation $\tilde{\phi}_i = \phi_i(\tilde x)$ holds. The question of how to operationally relate values of different fields, and assign and distribute them into $n$-tuples, is a matter of a separate study \cite{JanjicPaunkovicVojinovic2020}. In this work, we assume that this problem is already solved. Moreover, note that fields $\phi$ need not be observable, due to potential gauge symmetries (for example, the electromagnetic potential $A_{\mu}$ and the metric $g_{\mu\nu}$). To that end, we introduce an ordered set of gauge invariant functions $\cO(\phi) \equiv (\cO_1(\phi),\dots, \cO_m(\phi))$, where $m\geq n$ (for example, the electromagnetic field strength $F_{\mu\nu}$ and the curvature $R^{\lambda}{}_{\mu\nu\rho}$), and {\em define a spacetime point} as an $m$-tuple of their values $\tilde\cO$.

Unless the physical system features some global symmetry, each $m$-tuple $\tilde\cO$ defines a unique point in spacetime. Note that, in the context of GR, the absence of global symmetries is actually the generic case. Thus, the essential feature of this definition is that it does not make sense to say that the same $m$-tuple of field strengths can occur in two different spacetime points, since ``both'' spacetime points in question are defined in terms of the one and the same $m$-tuple, and therefore represent a {\em single} point.

Moving from classical to the quantum framework, where no system has predetermined physical properties independent of observation, one needs to talk about observables. Given an ordered set of quantum fields $\phi \equiv (\phi_1,\dots,\phi_n )$, one constructs one {\em specific} complete set of compatible observables $\cO \equiv (\cO_1(\phi), \dots. \cO_m(\phi))$, where
\begin{itemize}
\item {\em compatible} means that all observables mutually commute, $\komut{\cO_i}{\cO_j} = 0$ for every $i$ and $j$, while
\item {\em complete} means that the eigenspaces common for all these observables are nondegenerate, i.e., they are one-dimensional subspaces of the total Hilbert space.
\end{itemize}
Here, by ``specific'' we mean the set of observables which depend {\em only} on fields $\phi$, but {\em not} on their conjugated momenta. This fixes the coordinate representation, such that each common eigenvector corresponds to one classical configuration of fields. The outcomes of the measurements of these observables can then be grouped into $m$-tuples and used to {\em define individual spacetime points}, as in the classical case above, thus giving rise to an emergent classical spacetime. On the other hand, if the state is not an eigenvector of $\cO$, one cannot speak of a single classical configuration of fields, and thus the notion of emergent spacetime and its points ceases to make sense globally, according to the relational approach. At most, one could speak of a superposition of classical configurations and corresponding emergent spacetimes, but without any natural way to relate spacetime points across different branches in the superposition. Nevertheless, this does not mean that establishing such a relation is impossible for certain subregions of spacetime. Indeed, the whole non-yellow ``outside'' part of the gravitational switch picture from Subsection \ref{sec:gravitationalswitch} represents a subregion with a locally classical configuration and thus well defined spacetime points.

In order to better appreciate the relational definition of spacetime points given above, it is instructive to look at the realisation of spacetime in the context of a relational quantum gravity model, such as a spinfoam model in the Loop Quantum Gravity (LQG) framework~\cite{rov:04,rov:vid:14}. There, the spacetime is ``built'' out of the spin foam --- a lattice-like structure with vertices, edges and faces, each labeled by the eigenvalues of particular field operators that ``live'' on these structures, depicted as follows \blue{(see Figure~\ref{sl:deset})}:

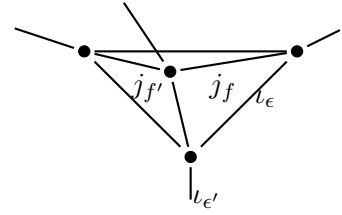
\begin{figure}[!ht]
\begin{center}
\begin{tikzpicture}[scale=0.7]
\node at (0,0,0) (anode) {};
\node at (0,2,1) (bnode) {};
\node at (-2,2,0) (cnode) {};
\node at (2,2,0) (dnode) {};

\node at (0,-1,0) (apnode) {};
\node at (-1,3.5,1) (bpnode) {};
\node at (-3.5,2.5,0) (cpnode) {};
\node at (3,2.5,0) (dpnode) {};

\filldraw[black] (anode) circle (3pt) {};
\filldraw[black] (bnode) circle (3pt) {};
\filldraw[black] (cnode) circle (3pt) {};
\filldraw[black] (dnode) circle (3pt) {};

\draw[thick] (anode) to (bnode);
\draw[thick] (anode) to (cnode);
\draw[thick] (anode) to (dnode);
\draw[thick] (bnode) to (cnode);
\draw[thick] (bnode) to (dnode);
\draw[thick] (cnode) to (dnode);

\draw[thick] (anode) to (apnode);
\draw[thick] (bnode) to (bpnode);
\draw[thick] (cnode) to (cpnode);
\draw[thick] (dnode) to (dpnode);

\node at (0.8,1.5,0.5) {$j_f$};
\node at (-0.6,1.5,0.5) {$j_{f^\prime}$};
\node at (1.4,1.1,0) {$\iota_{\epsilon}$};
\node at (0.3,-0.8,0) {$\iota_{\epsilon^\prime}$};

\end{tikzpicture}
\end{center}
\caption{\blue{A piece of a spin foam diagram. The field $j$ labels the faces $f, f',\dots$, while the field $\iota$ labels the edges $\epsilon,\epsilon',\dots$, of the diagram.}}
\label{sl:deset}
\end{figure}
For example, the {\em area operator}, which is a function of the gravitational field, has eigenvalues determined by a half-integer label $j\in \prirodni/2$, and each face of the spin foam carries one such label, specifying the area of the surface dual to that face. In particular, the spectrum of the area operator is given as
\begin{equation}
A(j) = 8\pi \gamma l_p^2 \sum_{f} \sqrt{j_f(j_f+1)} \,,
\end{equation}
where $l_p$ is the Planck length, $\gamma$ is the Barbero-Immirzi parameter, while the sum goes over all faces $f$ of the spin foam that intersect the surface whose area we are interested in, see~\cite{rov:04,rov:vid:14} for details. All other physical observables similarly provide appropriate labels for each vertex, edge and face of the spin foam. Since edges and faces meet at vertices, a given vertex carries labels of all observables of all edges and faces that are connected to that vertex. These observables form the complete set of compatible observables $\cO$, and their eigenvalues label each vertex, determining the identity of that vertex. In other words, each labeled vertex of a spin foam defines a ``spacetime point'', and if two vertices have completely identical properties in the sense of their labels and their connectedness to neighbouring objects, they actually represent the one and the same vertex.

At first sight, it is tempting to apply the ideas of relational spacetime to the case of the quantum switch, as follows. At the spacetime event $A$, Alice interacts with the particle as it enters and exits her lab, while at the spacetime event $A^\prime$ Alice also interacts (in exactly the same way) with the same particle. The idea of relational spacetime then might suggest that one should {\em define} the spacetime events $A$ and $A^\prime$ by the physical event of interaction between Alice and the particle. Since this interaction is the same in both cases, one ought to identify the two points, $A\equiv A^\prime$, and claim that both of these correspond to the same spacetime event, defined by the interaction between Alice and the particle. The same argument applies to Bob, and events $B$ and $B^\prime$.

Unfortunately, this argument is \blue{not fully in line with relationalism.} The reason lies in the fact that the interaction between Alice and the particle (and also between Bob and the particle) does not meet the criteria given in the above relational definition of a spacetime point. Namely, neither Alice, nor Bob, performs a measurement of a {\em complete set of compatible observables} $\cO$. The mentioned interaction with the particle is merely a subset of this. In particular, the interaction of Alice with the particle does not uniquely fix the state of, say, the gravitational field, or the electromagnetic field, or the Higgs field, etc. Therefore, it may happen that the measurement outcomes of the whole set of observables $\cO$ at spacetime events $A$ and $A^\prime$ are still mutually distinct, thereby defining the events $A$ and $A^\prime$ as two distinguishable spacetime points. In order to be certain that $A$ and $A^\prime$ are really the same spacetime event, Alice would need to measure the complete set of observables $\cO$, and convince herself that the results of all those measurements at $A$ and at $A^\prime$ are identical. The mere interaction with the particle is not enough to achieve this, and the experimental setups such as~\cite{pro:etal:15,rub:roz:fei:ara:zeu:pro:bru:wal:17,rub:roz:mas:ara:zyc:bru:wal:17} obviously fall short of accounting for the state of all other possible physical fields that Alice and Bob can interact with, in addition to the interaction with the particle.

We see that, when applied to the case of the quantum switch in classical \purple{gravitational field}, the relational framework is at odds with the operational approach --- the former distinguishes $A$ and $A^\prime$ while the latter regards them as identical. \purple{This is because the matter fields of the particle are in a superposition of two classical configurations. Similarly, in the case of} the $2$-event gravitational switch introduced in Subsection \ref{sec:gravitationalswitch}, the overall state \purple{of gravity and matter} is a superposition of two distinct classical configurations. Therefore, within the relational framework, it is not possible to talk about a single emergent spacetime, nor to compare the points that belong to different branches. This is different from the operational approach, which aims to identify points from different branches. It is also different from the traditional approach, since the latter postulates a unique classical spacetime manifold.

Note that, if understood as an {\em interpretation}, relational framework ought to have all experimental predictions the same as those from the traditional approach. Thus, the observable constructed in Subsection \ref{sec:distinguishing_without_decohering} should distinguish the quantum from the gravitational switch, in the same way as in the traditional approach. On the other hand, potential new physics formulated based on the relational framework might, or might not, feature different experimental predictions.

It is important to emphasise that, as discussed in Subsection \ref{sec:otherswitches}, various realisations of the quantum switch are possible by superposing different causal orders in the framework of QG. In particular, regarding the $2$-event realisations, one can consider the following diagram \blue{(see Figure~\ref{sl:jedanaest})}:

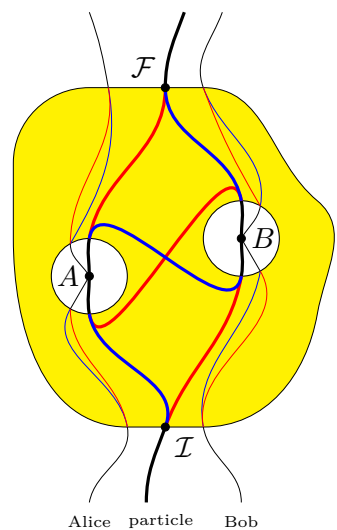
\begin{figure}[h!]
\begin{center}
\begin{tikzpicture}[scale=0.5]

\draw[very thin, fill=yellow] (5,2) to [out=0, in=180 ] (7,2) to [out=0, in=260] (10,5) to [out=80, in=320] (10,8) to [out=140, in=0] (7,11) to [out=180, in=0] (4,11) to [out=180, in=90] (2,9) to [out=270, in=90] (2,7) to [out=270, in=180] (5,2);

\filldraw [fill=white] (4,6) circle (1);
\filldraw [fill=white] (8,7) circle (1);

\node at (5.5,-0.5) {$\quad$\tiny particle};
\draw[very thick, black] (5.5,0) to [out=90 , in=250] (6,2);
\draw[very thick, black] (6,11) to [out=90 , in=250] (6.5,13);

\draw[very thick, black] (8,6) to [out=80 , in=270] (8,7) to [out=90 , in=280] (8,8);
\draw[very thick, black] (4,5) to [out=100 , in=270] (4,6) to [out=90 , in=260] (4,7);

\draw[very thick, red] (6,2) to [out=70, in=260] (8,6);
\draw[very thick, red] (8,8) to [out=100 , in=280] (4,5);
\draw[very thick, red] (4,7) to [out=80 , in=270] (6,11); 

\draw[very thick, blue] (6,2) to [out=70, in=280] (4,5);
\draw[very thick, blue] (4,7) to [out=80 , in=260] (8,6);
\draw[very thick, blue] (8,8) to [out=100 , in=270] (6,11); 

\node at (4,-0.5) {\tiny Alice};
\draw[very thin, black] (4,0) to [out=80, in=280] (5,2);
\draw[very thin, black] (4.5,11) to [out=100, in=290] (4,13);
\draw[very thin, black] (3.5,5.13) to [out=60, in=240] (4,6) to [out=120, in=270] (3.5,6.86);

\draw[very thin, blue] (5,2) to [out=100, in=240] (3.5,5.13);
\draw[very thin, blue] (3.5,6.86) to [out=60, in=280] (4.5,11);

\draw[very thin, red] (5,2) to [out=100, in=270] (3.5,5.13);
\draw[very thin, red] (3.5,6.86) to [out=90, in=280] (4.5,11);

\node at (8,-0.5) {\tiny Bob};
\draw[very thin, black] (8,0) to [out=90, in=280] (7,2);
\draw[very thin, black] (7,11) to [out=110, in=250] (7.5,13);
\draw[very thin, black] (8.5,6.13) to [out=120, in=300] (8,7) to [out=60, in=260] (8.5,7.86);

\draw[very thin, blue] (7,2) to [out=100, in=260] (8.5,6.13);
\draw[very thin, blue] (8.5,7.86) to [out=80, in=290] (7,11);

\draw[very thin, red] (7,2) to [out=100, in=300] (8.5,6.13);
\draw[very thin, red] (8.5,7.86) to [out=120, in=290] (7,11);

\filldraw[black] (4,6) circle (3pt) node[anchor=east] {$A$};
\filldraw[black] (8,7) circle (3pt) node[anchor=west] {$B$};
\filldraw[black] (6,2) circle (3pt) node[anchor=north west] {$\mathcal{I}$};
\filldraw[black] (6,11) circle (3pt) node[anchor=south east] {$\mathcal{F}$};

\end{tikzpicture}
\end{center}
\caption{\blue{Spacetime diagram of a version of a $2$-event gravitational switch, in which Alice and Bob perform their respective operations in the regions of spacetime with a single gravitational configuration.}}
\label{sl:jedanaest}
\end{figure}
This diagram features two {\em classical} spacetime subregions surrounding Alice's and Bob's laboratories. As such, Alice and Bob can measure the complete set of compatible observables within their laboratories, without obtaining which-way information and destroying the superposition. Therefore, even from the relational point of view, this represents an implementation of a $2$-event gravitational switch. Note that in this case Alice and Bob do not even need Friend in order to verify the $2$-event nature of their gravitational switch.

It is interesting to observe that this realisation of the quantum switch implements the operational idea of a $2$-event quantum switch, in terms of closed laboratories. However, to achieve such an implementation, it is necessary to have a genuine superposition of metric-induced \blue{spacetime} causal orders in the yellow region of spacetime, which does not feature in experimental realisations \cite{pro:etal:15,rub:roz:fei:ara:zeu:pro:bru:wal:17,rub:roz:mas:ara:zyc:bru:wal:17}.

\section{\label{sec:conclusions}Conclusions}

 
In this paper, we analysed the notion of causal orders both in classical and quantum worlds, with the emphasis on the latter. We defined the notion of the causal order for the case of (classical and quantum) circuits, in terms of partial ordering between the nodes of the circuit's underlying graph that defines the cause-effect structure. We discussed the possibility of implementing an abstract circuit in the real world, showing that it is always possible to do so for the case of a globally hyperbolic (classical) spacetime, in which the circuit's causal order is preserved by the metric-induced relation between the spacetime events.

The superposition principle of quantum mechanics offers the possibility of controlled operations, in particular the quantum switch, whose experimental realisations have been claimed to present genuine superpositions of causal orders. Within the process matrix formalism, we have analysed the $4$- and $3$-event realisations of the quantum switch in classical spacetimes with fixed \blue{spacetime} causal orders, and the 2-event realisation of a gravitational switch that features superpositions of different gravitational field configurations and their respective \blue{spacetime} causal orders. To that end, we have extended the process matrix formalism, by introducing the notion of a vacuum state. Our analysis shows that the process matrix formalism can explain the quantum switch realisations within the standard physics, and is thus consistent with it.

Thus, as a consequence of our Theorem, and the analysis of the quantum switch implementations, we argued that, \blue{in contrast to the gravitational switch,} the current experimental implementations do not feature superpositions of \blue{spacetime} causal orders, and that they are variants of the time double slit experiment. Moreover, by explicitly constructing two different observables, presented in Sections~\ref{sec:distinguishing_decohering} and~\ref{sec:distinguishing_without_decohering}, respectively, we showed that it is possible to experimentally distinguish between different realisations of the quantum switch.

Finally, in Section~\ref{sec:discussion}, we analysed the relation among the traditional QFT approach to QG (used throughout this paper), the operational point of view, and the relational framework of QG. On the example of the quantum switch, we showed that the operational viewpoint, while consistent with the approach advocated by Mach, is nevertheless at odds with the modern relational framework. On the other hand, the traditional QFT approach and the relational framework may or may not be compatible, depending on the concrete realisation of the quantum switch. In particular, for the specific realisation of the gravitational switch given in Subsection \ref{sec:relationalism}, the two frameworks are compatible in the prediction that Alice and Bob can locally (without the help of Friend) verify that the switch is implemented on $2$ events. 


In a recent work~\cite{ho:cos:gia:ral:18}, the authors report on a violation of the causal inequality~\cite{ore:cos:bru:12} in flat Minkowski spacetime with a definite causal order. To achieve it, they consider laboratories that are localised in space only, while delocalised in time. Therefore, their alternative notion of a ``closed laboratory'', and that considered in~\cite{ore:cos:bru:12}, do not coincide, this way manifestly violating the conditions necessary for the causal inequality to hold. For the same reason, the scenario considered in~\cite{ho:cos:gia:ral:18} falls out of the scope of the current work as well. \blue{Additionally, in another recent work~\cite{ore:19}, the author discusses the quantum switch in terms of the time-delocalised quantum subsystems and operations, and generalises it to more complex quantum circuits and processes. The results of these two papers deserve further analysis and remain to be a subject of future research.}


Exploring possible generalisations of our Theorem, as suggested at the end of Appendix~\ref{sec:AppProof}, presents a straightforward future line of research. Also, one could further analyse the process matrix formalism, in particular by exploring the situations in which the operational approach interpretation fails to describe the known processes. Or, to search for the opposite --- the instances of physical processes that cannot be explained by the process matrix formalism, when applied within the standard physics. In order to show that the process matrix formalism is perfectly suitable for describing the quantum switch implementations within the standard physics, we formulated its version that features the vacuum state.  One can thus further study possible generalisations of this formalism and its applications to the cases that go beyond simple non-relativistic mechanics. Finally, motivated by our analysis and discussion from Subsections \ref{sec:distinguishing_decohering} and \ref{sec:mach}, one can try to formulate alternative theories that would be consistent with the experimentally tested known physics (GR in particular), while at the same time \blue{substituting the spacetime events $A$ and $A^\prime$}, from the quantum switch realisations in classical spacetimes\blue{, with a single operational event (and analogously for $B$ and $B^\prime$).}

\bigskip

\centerline{\bf Acknowledgments}

\bigskip

The authors would like to thank \v Caslav Brukner, Esteban Castro-Ruiz and Flaminia Giacomini for helpful comments and discussions.

NP acknowledges the support of SQIG --- Security and Quantum Information Group, the Instituto de Telecomunica\c{c}\~oes (IT) Research Unit, ref. UIDB/50008/2020, funded by Funda\c{c}\~ao para a Ci\^{e}ncia e Tecnologia (FCT), and the FCT projects QuantumMining POCI-01-0145-FEDER-031826 and Predict PTDC/CCI-CIF/29877/2017, supported by the European Regional Development Fund (FEDER), through the Competitiveness and Internationalisation Operational Programme (COMPETE 2020), and by the Regional Operational Program of Lisbon, as well as the FCT Est\'{i}mulo ao Emprego Cient\'{i}fico grant no. CEECIND/04594/2017/CP1393/CT0006. The bilateral scientific cooperation between Portugal and Serbia through the project ``Noise and measurement errors in multi-party quantum security protocols'', no. 451-03-01765/2014-09/04 supported by the FCT, Portugal, and the Ministry of Education, Science and Technological Development of the Republic of Serbia is also acknowledged.

MV was supported by the project ON~171031 of the Ministry of Education, Science and Technological Development of the Republic of Serbia, and the bilateral scientific cooperation between Austria and Serbia through the project ``Causality in Quantum Mechanics and Quantum Gravity --- 2018-2019'', no.  451-03-02141/2017-09/02 supported by the Austrian Academy of Sciences (\"OAW), Austria, and the Ministry of Education, Science and Technological Development of the Republic of Serbia.

\bibliographystyle{plain}

\onecolumn\newpage
\appendix

\section{\label{sec:AppProof}Proof of the Theorem}

Here we give an explicit constructive proof of the Theorem from the main text.

\medskip

Given the graph $G$, we begin the proof by partitioning its set of nodes $I$ into disjoint subsets, in the following way. Since the graph is finite, we introduce the subset $M_1 \subset I$ which consists of all minimal nodes of the graph G:
\begin{equation}
\label{eq:minimal_nodes}
M_1 = \{ u\in I \, | \, (\neg \exists v\in I)\, v \orderi u \} \,.
\end{equation}
Since all nodes in $M_1$ are minimal, there is no order relation $\orderi$ between any two of them. Therefore, we can intuitively understand them as ``simultaneous''. As a next step, we remove these nodes and the corresponding edges from $G$, reducing it to a subgraph $G_2 = (I_2,E_2)$, where
\begin{equation}
I_2 = I \backslash N_1\,, \qquad E_2 = \{ (u,v) \,| \, u,v\in I_2\,, \, (u,v)\in E \}\,.
\end{equation}
Then we repeat the construction for the graph $G_2$, obtaining the new minimal set $M_2$, and the next subgraph $G_3$, in an analogous way. Since the graph $G$ is finite, after a certain finite number of steps we will exhaust all nodes in $I$, ending up with a partition of ``simultaneous'' subsets $M_1,\dots,M_m$ ($m\in\prirodni$), such that
\begin{equation}
(\forall i\neq j) \quad
M_i \cap M_j = \emptyset\,, \qquad \bigcup_{i=1}^m M_i = I\,.
\end{equation}

Once we have partitioned the set of nodes $I$ into subsets, we turn to the construction of the immersing map $P: I\to \cM$, in the following way. Since spacetime is globally hyperbolic, we can write $\cM = \Sigma \times \realni$, where $\Sigma$ is a spatial $3$-dimensional hypersurface, and $\realni$ is timelike. Without loss of generality, one can then introduce a foliation of spacetime into a family of such hypersurfaces, denoted $\Sigma_t$ and labeled by a parameter $t\in\realni$. Start from some initial parameter $t_1$, and choose a compact subset $S_{t_1} \subset \Sigma_{t_1}$. Denoting the number of elements in the partition $M_i$ as $\Vert M_i \Vert$, we pick in an arbitrary way the set of $\Vert M_1 \Vert$ points $\vec{x}_k\in S_{t_1}$ (here, $k=1,\dots,\Vert M_1 \Vert$), and define the map $P$ to assign a node from $M_1$ to each point $\vec{x}_k$ in a one-to-one fashion:
\begin{equation}
P(u_k) = (t_1,\vec{x}_k) \in \cM\,, \qquad k=1,\dots,\Vert M_1 \Vert\,.
\end{equation}
Once this assignment has been defined, construct a future-pointing light cone from each spacetime point $(t_1,\vec{x}_k)$. Then we find a new hypersurface, $\Sigma_{t_2}$, which contains a common intersection with all constructed light cones, and denote this intersection $S_{t_2} \subset \Sigma_{t_2}$. In this way, by construction, all points $(t_1,\vec{x}_k)$ are in the past of all points in $S_{t_2}$,
\begin{equation}
(t_1,\vec{x}_k) \orderm S_{t_2}\,, \qquad k=1,\dots,\Vert M_1 \Vert\,.
\end{equation}
Now extend the definition of $P$ such that it assigns the nodes from the next partition, $M_2$, to a randomly chosen set of points in $S_{t_2}$ in a similar way as before, then construct a set of light cones from them, and repeat the construction for all partitions $M_i$. Constructed in this way, the map $P$ ensures that for every pair of nodes $u,v \in I$, we have
\begin{equation} \label{eq:ImmersingRequirementOnNodes}
u \orderi v \quad \Longrightarrow \quad P(u) \orderm P(v)\,, \qquad \forall u,v \in I\,.
\end{equation}
Once we have constructed the map $P: I\to \cM$ satisfying (\ref{eq:ImmersingRequirementOnNodes}), using the definition (\ref{eq:giOrderEquivalence}), it induces the map $\cP : \cG_\cC \to \cM$, which satisfies the required statement (\ref{eq:ImmersingRequirement}).

This completes the proof. \hfill $\Box$

\medskip

Note that, while the causal order $\orderm$ indeed preserves the causal order $\orderc$, it is ``stronger'' in the sense that it may introduce additional relations between the images of nodes, which do not hold in the graph itself. Indeed, the construction of the map $P$ in the above proof is such that {\em each} image of a node from some given partition $M_i$ is in the causal past of {\em all} images from the previous partition $M_{i-1}$, which is not necessarily the case for the nodes themselves. One might study if the causal orders over the set of nodes and over the set of its images can be equivalent, i.e., if the opposite implication from equation (\ref{eq:ImmersingRequirementOnNodes}) also holds (in this case the immersion $P$ is called an {\em embedding} of $G$ into $\cM$). Whether such an embedding exists for all hyperbolic spacetimes, or at least for some, is an open question.

Next, one could also discuss the generalisation of the above theorem to the case of countably infinite graphs $G$. However, for our purposes, the existence of a partially ordered map $\cP$ over the set of finite graphs will suffice.

Regarding the proof itself, one can formulate an alternative (and simpler) approach to the proof of the theorem. Namely, one can first prove that every circuit can be immersed into the flat Minkowski spacetime. Then, knowing that the sufficiently small neighbourhood of every spacetime point in an arbitrary manifold $\cM$ can be well approximated with its tangent space, one can always immerse the whole circuit into this small neighbourhood. However, this implies that the geometric size of the circuit can be considered negligible compared to the curvature scale of the manifold, which may render such implementation practically unfeasible. Moreover, this alternative approach does not cover the cases where one actually wants the scale of the circuit to be comparable to the curvature scale. Specifically, if one wishes to employ the circuit to study gravitational phenomena, its gates must be distributed across spacetime precisely in a way that is sensitive to curvature. Therefore, the construction of the map $P$ used in the proof of the theorem is more general than the construction in this alternative approach.

\begin{figure}[h!]
\begin{center}
\begin{tikzpicture}[scale=0.75]

\draw[->] (-0.5,0) -- (-0.5,6);
\node[anchor=east] at (-0.5,6) {time};

\node at (1,1) (ione) {};
\node at (2,1) (itwo) {};
\node at (3.5,1) (idots) {};
\node at (5,1) (iM) {};
\filldraw[black] (ione) circle (3pt) node[anchor=north] {$i_1\strut$};
\filldraw[black] (itwo) circle (3pt) node[anchor=north] {$i_2\strut$};
\filldraw[black] (idots) circle (3pt) node[anchor=north] {$\vphantom{i_3}\dots\strut$};
\filldraw[black] (iM) circle (3pt) node[anchor=north] {$i_{\|M_1\|}$};
\draw (0,0.3) rectangle (6,1.5);
\node[anchor=west] at (6,0.9) {$\cI\strut$};

\node at (1,5) (fone) {};
\node at (2,5) (ftwo) {};
\node at (3.5,5) (fdots) {};
\node at (5,5) (fM) {};
\filldraw[black] (fone) circle (3pt) node[anchor=south] {$f_1\strut$};
\filldraw[black] (ftwo) circle (3pt) node[anchor=south] {$f_2\strut$};
\filldraw[black] (fdots) circle (3pt) node[anchor=south] {$\vphantom{f_3}\dots\strut$};
\filldraw[black] (fM) circle (3pt) node[anchor=south] {$f_{\|M_m\|}$};
\draw (0,4.5) rectangle (6,5.7);
\node[anchor=west] at (6,5.1) {$\cF\strut$};

\node at (1.5,2) (one) {};
\node at (2.5,2) (two) {};
\node at (4.5,2) (three) {};
\node at (0.5,4) (four) {};
\node at (5.5,4) (six) {};
\node at (1.5,4) (seven) {};
\node at (3,4) (eight) {};
\node at (4,4) (nine) {};
\node at (2,3) (dotsone) {$\cdots$};
\node at (4,3) (dotstwo) {$\cdots$};
\filldraw[black] (one) circle (3pt);
\filldraw[black] (two) circle (3pt);
\filldraw[black] (three) circle (3pt);
\filldraw[black] (four) circle (3pt);
\filldraw[black] (six) circle (3pt);
\filldraw[black] (seven) circle (3pt);
\filldraw[black] (eight) circle (3pt);
\filldraw[black] (nine) circle (3pt);
\draw[dashed] (0,1.8) rectangle (6,4.2);
\node[anchor=west] at (6,3) {$\cO_\cC \strut$};

\draw[->] (ione) -- (one);
\draw[->] (itwo) -- (one);
\draw[->] (idots) -- (two);
\draw[->] (idots) -- (three);
\draw[->] (iM) -- (three);
\draw[->] (one) -- (dotsone);
\draw[->] (two) -- (dotsone);
\draw[->] (two) -- (dotstwo);
\draw[->] (three) -- (dotstwo);

\draw[->] (dotsone) -- (four);
\draw[->] (dotsone) -- (seven);
\draw[->] (dotsone) -- (eight);

\draw[->] (dotstwo) -- (eight);
\draw[->] (dotstwo) -- (nine);
\draw[->] (dotstwo) -- (six);

\draw[->] (four) -- (fone);
\draw[->] (seven) -- (fone);
\draw[->] (seven) -- (ftwo);
\draw[->] (eight) -- (fdots);
\draw[->] (nine) -- (fdots);
\draw[->] (nine) -- (fM);
\draw[->] (six) -- (fM);

\end{tikzpicture}
\end{center}
\caption{\blue{The spacetime diagram of the circuit $\cC$, with the initial gate $\cI$, the operation gate $\cO_{\cC}$, and the final gate~$\cF$.}}
\label{sl:dvanaest}
\end{figure}
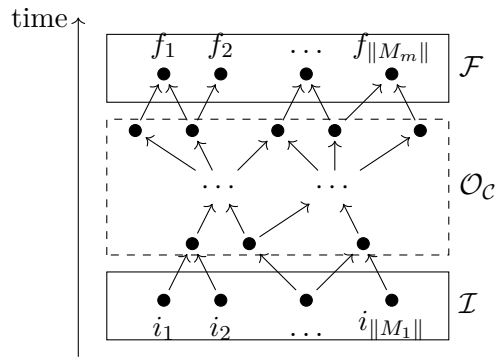

Finally, given the construction in the proof, the gates of the set of minimal nodes $M_1$ define the initial gate $\cI$, the set of maximal nodes $M_m$ define the final gate $\cF$, while the the gates of the remaining intermediary sets of nodes $M_2, \dots M_{m-1}$ define the operation $\cO_\cC$. This is illustrated in the diagram below \blue{(see Figure~\ref{sl:dvanaest})}.

\section{\label{sec:AppI}Qutrit states, operators and bases}

The notion of a qubit can be generalised from a $2$-dimensional Hilbert space to a $d$-dimensional Hilbert space. The generalised object is called ``qudit'' in $d$ dimensions~\cite{ber:kra:08}. Since we are interested in describing ordinary $2$-dimensional qubits with an additional vacuum state, it is natural to consider qudits in $d=3$, called ``qutrits''. We introduce the following notation for the basis states of a qutrit in $\cH_3=\kompleksni^3$:
\begin{equation} \label{eq:qutricComputationalBasis}
\ket{0} \equiv \left[
\begin{array}{c}
1 \\ 0 \\ 0 \\
\end{array}
\right]\,, \qquad
\ket{1} \equiv \left[
\begin{array}{c}
0 \\ 1 \\ 0 \\
\end{array}
\right]\,, \qquad
\ket{v} \equiv \left[
\begin{array}{c}
0 \\ 0 \\ 1 \\
\end{array}
\right]\,.
\end{equation}
The states $\ket{0}$ and $\ket{1}$ will be understood as the usual computational basis for a $2$-dimensional qubit, while the state $\ket{v}$ will represent the vacuum, i.e., the ``absence of a qubit''. In cases when we take sums over the basis vectors, we will assume that the vacuum state carries the index $2$, i.e., $\ket{v} \equiv \ket{2}$, so that we can write
\begin{equation}
\sum_{i=0}^2 \ket{i} = \ket{0} + \ket{1} + \ket{v}\,, \qquad \text{ and } \qquad \sum_{i=0}^1 \ket{i} = \ket{0} + \ket{1}\,.
\end{equation}
Using this notation, we write the unnormalised maximally correlated states for the qutrit and the qubit as
\begin{equation} \label{eq:TransportVectorsDef}
\kket{\one} = \sum_{i=0}^2 \ket{i} \ket{i} = \ket{0}\ket{0} + \ket{1}\ket{1} + \ket{v}\ket{v} \in \cH_3 \otimes \cH_3\,, \qquad
\kket{1} = \sum_{i=0}^1 \ket{i} \ket{i} = \ket{0}\ket{0} + \ket{1}\ket{1} \in \cH_2 \otimes \cH_2\,,
\end{equation}
so that
\begin{equation} \label{eq:TransportVectorsRelation}
\kket{\one} = \kket{1} + \ket{v}\ket{v}\,.
\end{equation}

One can also introduce the standard Hilbert-Schmidt basis in the space $\cL(\cH_3)$ of linear operators on $\cH_3$. This basis consists of $9$ matrices $3\times 3$, labeled as $\lambda_0,\dots,\lambda_8$, as follows:
\begin{itemize}
\item the three symmetric matrices
\begin{equation}
\lambda_1 = \sqrt{\frac{3}{2}} \left[
\begin{array}{ccc}
 0 & 1 & 0 \\
 1 & 0 & 0 \\
 0 & 0 & 0 \\
\end{array}
  \right]\,,
\qquad
\lambda_2 = \sqrt{\frac{3}{2}} \left[
\begin{array}{ccc}
 0 & 0 & 1 \\
 0 & 0 & 0 \\
 1 & 0 & 0 \\
\end{array}
  \right]\,,
\qquad
\lambda_3 = \sqrt{\frac{3}{2}} \left[
\begin{array}{ccc}
 0 & 0 & 0 \\
 0 & 0 & 1 \\
 0 & 1 & 0 \\
\end{array}
  \right]\,,
\end{equation}
\item the three antisymmetric matrices
\begin{equation}
\lambda_4 = \sqrt{\frac{3}{2}} \left[
\begin{array}{ccc}
 0 & -i & 0 \\
 i &  0 & 0 \\
 0 &  0 & 0 \\
\end{array}
  \right]\,,
\qquad
\lambda_5 = \sqrt{\frac{3}{2}} \left[
\begin{array}{ccc}
 0 & 0 & -i \\
 0 & 0 &  0 \\
 i & 0 &  0 \\
\end{array}
  \right]\,,
\qquad
\lambda_6 = \sqrt{\frac{3}{2}} \left[
\begin{array}{ccc}
 0 & 0 &  0 \\
 0 & 0 & -i \\
 0 & i &  0 \\
\end{array}
  \right]\,,
\end{equation}
\item and the three diagonal matrices
\begin{equation}
\lambda_7 = \sqrt{\frac{3}{2}} \left[
\begin{array}{ccc}
 1 &  0 & 0 \\
 0 & -1 & 0 \\
 0 &  0 & 0 \\
\end{array}
  \right]\,,
\qquad
\lambda_8 = \frac{1}{\sqrt{2}} \left[
\begin{array}{ccc}
 1 & 0 &  0 \\
 0 & 1 &  0 \\
 0 & 0 & -2 \\
\end{array}
  \right]\,,
\qquad
\lambda_0 = \left[
\begin{array}{ccc}
 1 & 0 & 0 \\
 0 & 1 & 0 \\
 0 & 0 & 1 \\
\end{array}
  \right]\,.
\end{equation}
\end{itemize}
The matrix $\lambda_0$ is the unit matrix, while $\lambda_1,\dots,\lambda_8$ are self-adjoint, traceless, and orthogonal with respect to the standard scalar product:
\begin{equation}
\lambda_i^{\dag} = \lambda_i\,, \qquad \tr \lambda_i = 0\,, \qquad \tr \lambda_i^{\dag} \lambda_j = 3 \delta_{ij}\,, \qquad i=1,\dots,8\,.
\end{equation}
They represent the generators of the $SU(3)$ group, and are known as the Gell-Mann matrices (up to a normalisation factor $\sqrt{3/2}$).

If we denote $\cH_v$ as the $1$-dimensional vacuum-spanned subspace of $\cH_3$, one can see that $\cL(\cH_2)\oplus \cL(\cH_v)\subset \cL(\cH_3)$. In particular, if we denote the standard Pauli matrices as $\sigma_x,\sigma_y,\sigma_z$ and the unit $2\times 2$ matrix as $I_2$, they form the basis in $\cL(\cH_2)$, and the qubit basis can thus be constructed as
\begin{equation}
\sqrt{\frac{2}{3}}\lambda_1 = \left[
\begin{array}{cc|c}
\multicolumn{2}{c|}{\multirow{2}{*}{$\sigma_x$}} & 0 \\
   &   & 0 \\ \hline
 0 & 0 & 0 \\
\end{array}
 \right]\,, \qquad
\sqrt{\frac{2}{3}}\lambda_4 = \left[
\begin{array}{cc|c}
\multicolumn{2}{c|}{\multirow{2}{*}{$\sigma_y$}} & 0 \\
   &   & 0 \\ \hline
 0 & 0 & 0 \\
\end{array}
 \right]\,, \qquad
\sqrt{\frac{2}{3}}\lambda_7 = \left[
\begin{array}{cc|c}
\multicolumn{2}{c|}{\multirow{2}{*}{$\sigma_z$}} & 0 \\
   &   & 0 \\ \hline
 0 & 0 & 0 \\
\end{array}
 \right]\,,
\end{equation}
along with
\begin{equation}
\frac{2}{3}\lambda_0 + \frac{\sqrt{2}}{3} \lambda_8 = \left[
\begin{array}{cc|c}
\multicolumn{2}{c|}{\multirow{2}{*}{$I_2$}} & 0 \\
   &   & 0 \\ \hline
 0 & 0 & 0 \\
\end{array}
 \right]\,.
\end{equation}
Also, the vacuum space $\cL(\cH_v)$ is one-dimensional, and the basis is
\begin{equation}
\frac{1}{3}\lambda_0 - \frac{\sqrt{2}}{3} \lambda_8 = \left[
\begin{array}{cc|c}
 0 & 0 & 0 \\
 0 & 0 & 0 \\ \hline
 0 & 0 & 1 \\
\end{array}
 \right]\,.
\end{equation}

\section{\label{sec:AppFourEvent}Process matrix evaluation}

Let us give an explicit step by step evaluation of the probability distribution for the $4$-event process discussed in the text, using the process matrix formalism. The complete spacetime diagram of the process is given as \blue{(see Figure~\ref{sl:trinaest})}:

\begin{figure}[h!]
\begin{center}
\begin{tikzpicture}[x={(0cm,1cm)},y={(1cm,0cm)},scale=0.8]

\draw[->] (3.5,-1.5) -- (9,-1.5);
\node[anchor=east] at (9,-1.5) {space};

\draw[->] (4,-2) -- (4,14);
\node[anchor=north] at (4,14) {time};
\draw[very thin] (3.9,0.5) -- (4.1,0.5);
\node[anchor=north] at (3.9,0.5) {$t_i$};
\draw[very thin] (3.9,3) -- (4.1,3);
\node[anchor=north] at (3.9,3) {$t_1$};
\draw[very thin] (3.9,8) -- (4.1,8);
\node[anchor=north] at (3.9,8) {$t_2$};
\draw[very thin] (3.9,12) -- (4.1,12);
\node[anchor=north] at (3.9,12) {$t_f$};

\draw[very thin] (5,-1.7) -- (5,9);
\node[anchor=east] at (5,-1.7) {\tiny Alice};

\draw[very thin] (8,-1.7) -- (8,9);
\node[anchor=east] at (8,-1.7) {\tiny Bob};

\node at (6.5,0.5) (splitterSi) {};
\node at (6.5,12) (splitterSf) {};
\draw[very thin] (6.5,-1.7) -- (6.5,-1.2);
\node[anchor=east] at (6.5,-1.7) {$\substack{ \text{\tiny beam} \\ \text{\tiny splitter} }$};
\filldraw[black] (splitterSi) circle (3pt) node[anchor=north] {$S_i\strut$};
\filldraw[black] (splitterSf) circle (3pt) node[anchor=north] {$S_f\strut$};

\node at (5,3) (bluegateA) {};
\node at (8,8) (bluegateBprime) {};
\filldraw[blue] (bluegateA) circle (3pt) node[anchor=north] {$A\strut$};
\filldraw[blue] (bluegateBprime) circle (3pt) node[anchor=south] {$B^\prime\strut$};

\node at (8,3) (redgateB) {};
\node at (5,8) (redgateAprime) {};
\filldraw[red] (redgateB) circle (3pt) node[anchor=south] {$B\strut$};
\filldraw[red] (redgateAprime) circle (3pt) node[anchor=north] {$A^\prime\strut$};

\draw[thick,blue] (splitterSi) -- (bluegateA) -- (bluegateBprime) -- (splitterSf);

\draw[thick,red] (splitterSi) -- (redgateB) -- (redgateAprime) -- (splitterSf);


\node at (6,-0.5) (gatePa) {};
\node at (7,-0.5) (gatePb) {};
\filldraw[black] (gatePa) circle (3pt) node[anchor=east] {$P_A\strut$};
\filldraw[black] (gatePb) circle (3pt) node[anchor=east] {$P_B\strut$};
\draw[thick] (gatePa) -- (splitterSi);
\draw[thick] (gatePb) -- (splitterSi);
\draw[dashed] (5.5,-1.3) rectangle (7.5,0.8);
\node[anchor=north] at (5.5,-0.25) {$\mathcal{I}$};

\node at (6,13.5) (gateTa) {};
\node at (7,13.5) (gateTb) {};
\filldraw[black] (gateTa) circle (3pt) node[anchor=west] {$T_A\strut$};
\filldraw[black] (gateTb) circle (3pt) node[anchor=west] {$T_B\strut$};
\draw[thick] (splitterSf) -- (gateTa);
\draw[thick] (splitterSf) -- (gateTb);
\draw[dashed] (5.5,11.7) rectangle (7.5,14.3);
\node[anchor=north] at (5.5,13) {$\mathcal{F}$};

\end{tikzpicture}
\end{center}
\caption{\blue{Spacetime diagram of the $4$-event implementation of the quantum switch. The internal structures of the composite gates $\cI$ and $\cF$ are explicitly depicted.}}
\label{sl:trinaest}
\end{figure}
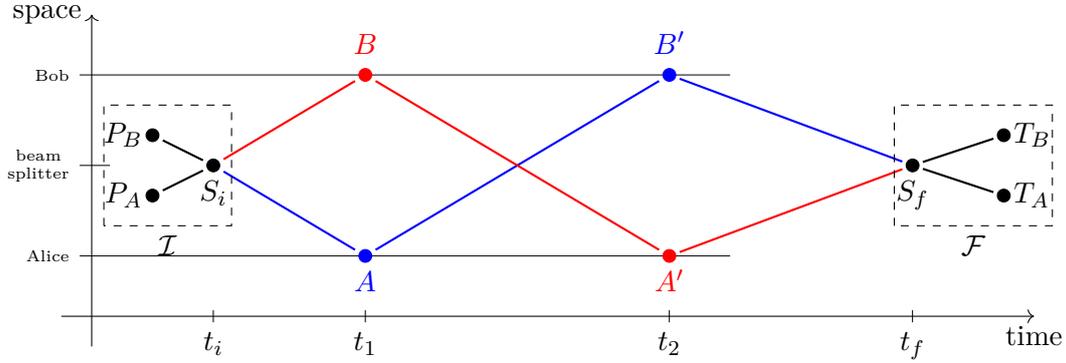

The composite event $\mathcal{I}$ consists of the two preparation events $P_A$, $P_B$, and the initial beam splitting event $S_i$, while $\mathcal{F}$ consists of the recombination event $S_f$ and the measurement events $T_A$ and $T_B$.

The corresponding circuit diagram is obtained from the above one by promoting each event of interaction to a gate, and the propagation of each particle to a channel.
This leads to the following circuit diagram \blue{(see Figure~\ref{sl:cetrnaest})}:

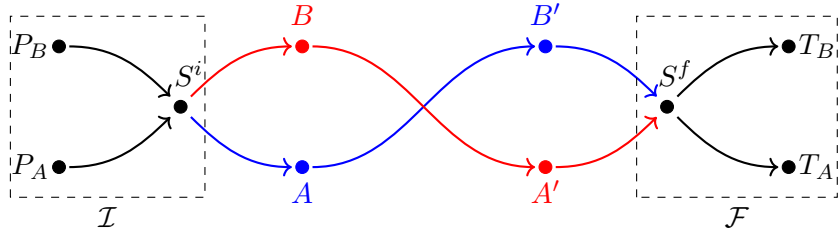
\begin{figure}[h!]
\begin{center}
\begin{tikzpicture}[x={(0cm,1cm)},y={(1cm,0cm)},scale=0.8]
\node at (0,-2) (gatePsi) {};
\node at (2,-2) (gateV) {};
\node at (1,0) (gateSi) {};
\node at (0,2) (gateA) {};
\node at (2,2) (gateB) {};
\node at (0,6) (gateAprime) {};
\node at (2,6) (gateBprime) {};
\node at (1,8) (gateSf) {};
\node at (0,10) (gateTA) {};
\node at (2,10) (gateTB) {};

\filldraw (0,-2) (gatePsi) circle (3pt) node[anchor=east] {$P_A\strut$};
\filldraw (2,-2) (gateV) circle (3pt) node[anchor=east] {$P_B\strut$};
\filldraw (1,0) (gateSi) circle (3pt) node[anchor=south] {$\;\; S^i\strut$};
\filldraw[blue] (0,2) (gateA) circle (3pt) node[anchor=north] {$A\strut$};
\filldraw[red] (2,2) (gateB) circle (3pt) node[anchor=south] {$B\strut$};
\filldraw[red] (0,6) (gateAprime) circle (3pt) node[anchor=north] {$A^\prime\strut$};
\filldraw[blue] (2,6) (gateBprime) circle (3pt) node[anchor=south] {$B^\prime\strut$};
\filldraw (1,8) (gateSf) circle (3pt) node[anchor=south] {$\;\; S^f\strut$};
\filldraw (0,10) (gateTA) circle (3pt) node[anchor=west] {$T_A\strut$};
\filldraw (2,10) (gateTB) circle (3pt) node[anchor=west] {$T_B\strut$};

\draw[->,thick] (gatePsi) to [out=0,in=-135] (gateSi);
\draw[->,thick] (gateV) to [out=0,in=135] (gateSi);
\draw[->,blue,thick] (gateSi) to [out=-45,in=180] (gateA);
\draw[->,blue,thick] (gateA) to [out=0,in=180] (gateBprime);
\draw[->,blue,thick] (gateBprime) to [out=0,in=135] (gateSf);
\draw[->,red,thick] (gateSi) to [out=45,in=180] (gateB);
\draw[->,red,thick] (gateB) to [out=0,in=180] (gateAprime);
\draw[->,red,thick] (gateAprime) to [out=0,in=-135] (gateSf);
\draw[->,thick] (gateSf) to [out=-45,in=180] (gateTA);
\draw[->,thick] (gateSf) to [out=45,in=180] (gateTB);

\draw[dashed] (-0.5,-2.8) rectangle (2.5,0.4);
\draw[dashed] (-0.5,7.5) rectangle (2.5,10.8);

\node[anchor=north] at (-0.5,-1.2) {$\mathcal{I}$};
\node[anchor=north] at (-0.5,9.15) {$\mathcal{F}$};

\end{tikzpicture}
\end{center}
\caption{\blue{Circuit diagram of the $4$-event implementation of the quantum switch. The internal structures of the composite gates $\cI$ and $\cF$ are explicitly depicted.}}
\label{sl:cetrnaest}
\end{figure}

Its structure is in one-to-one correspondence with the spacetime diagram for the $4$-event process, where the preparation and measurement spacetime events $\mathcal{I}$ and $\mathcal{F}$ have been split into three sub-gates each, for clarity.

The operations on each of the gates are given as follows. The preparation gate $P_A$ maps from the input Hilbert space $P_{A_I}$ to the output Hilbert space $P_{A_O}$, and analogously for gate $P_B$. The input spaces are trivial, $\dim P_{A_I} = \dim P_{B_I} = 1$, while each output space is spanned by vectors $\ket{0}$, $\ket{1}$ and $\ket{v}$. Here, $\ket{0}$ and $\ket{1}$ represent the two orthogonal qubit states (say, vertical and horizontal polarisations along certain axis in 3D space), while $\ket{v}$ is the vacuum state, representing the absence of particles in the corresponding arm of the interferometer. The operations performed at these gates, $P_A = \ket{\Psi}$ and $P_B = \ket{v}$, specify the initial conditions for the rest of the circuit diagram, and are described by the Choi-Jamio\l{}kowski (CJ) states as
\begin{equation} \label{eq:preparation_A}
\kket{P_A^\ast}{}^{P_{A_I}P_{A_O}} =  \ket{\Psi^\ast}{}^{P_{A_O}} \,, \qquad
\kket{P_B^\ast}{}^{P_{B_I}P_{B_O}} =  \ket{v}{}^{P_{B_O}} \,.
\end{equation}
Here, $\ast$ denotes the complex conjugation.

Analogously, the target gates $T_A$ and $T_B$ facilitate the final measurement outcomes of the circuit diagram. The input spaces $T_{A_I}$ and $T_{B_I}$ are three-dimensional, spanned over the two qubit states and the vacuum, while the output spaces are one-dimensional. The operations performed at these gates, $T_\alpha = \bra{\alpha}$ and $T_\beta = \bra{\beta}$, read out the measurement results $\alpha,\beta \in\{ 0,1,v \}$. The corresponding CJ states are given as
\begin{equation}
\label{eq:measurement_1}
\kket{T_{\alpha}^\ast}^{T_{A_I}T_{A_O}} =  \ket{\alpha}^{T_{A_I}} \, , \qquad
\kket{T_{\beta}^\ast}^{T_{B_I}T_{B_O}} =  \ket{\beta}^{T_{B_I}} \, .
\end{equation}

The gates $A$, $A^\prime$, $B$ and $B^\prime$ perform the unitaries $U$ and $V$. The input and output spaces $A_I$ and $A_O$ of the Alice's gate $A$ are both spanned by vectors $\ket{0}$, $\ket{1}$ and $\ket{v}$, and analogously for the input and output spaces of the remaining three gates. Assuming that in her (spatially) local laboratory Alice performs the unitary $U$ on the particle's internal degree of freedom, the induced operation between the three-dimensional spaces $A_I$ and $A_O$ that include the vacuum states is given by
\begin{equation} \label{eq:u_tilde_def}
	\tilde{U}^{A_OA_I} = U^{A_OA_I}P_{01}^{A_IA_I} + I^{A_OA_I}P_v^{A_IA_I} \, ,
\end{equation} 
where $P_{01}^{A_IA_I} = \ket{0}^{A_I}\bra{0}^{A_I} + \ket{1}^{A_I}\bra{1}^{A_I}$, $P_v^{A_IA_I} = \ket{v}^{A_I}\bra{v}^{A_I}$, and $I^{A_OA_I}$ represents the identity map between the Hilbert spaces $A_O$ and $A_I$. The analogous construction also holds for the gate $A^\prime$, so the respective CJ states for the gates $A$ and $A^\prime$ are then given by:
\begin{equation}
\label{eq:unitaries}
\begin{array}{rl}
	\kket{\tilde{U}^\ast}^{A_IA_O} & = \left[I^{A_IA_I} \otimes (\tilde{U}^\ast)^{A_OA_I}\right]\kket{\one}^{A_IA_I}\,, \vphantom{\ds \int} \\
	\kket{\tilde{U}^\ast}^{A_I^\prime A_O^\prime} & = \left[I^{A_I^\prime A_I^\prime} \otimes (\tilde{U}^\ast)^{A_O^\prime A_I^\prime}\right]\kket{\one}^{A_I^\prime A_I^\prime}\,.\vphantom{\ds \int}
\end{array}
\end{equation}
Here, the ``transport vector'' is given by (for details of the process matrix formalism for the case of three-dimensional spaces --- qutrits, see Appendix~\ref{sec:AppI}):
\begin{equation}
\kket{\one} = \ket{0}\ket{0} + \ket{1}\ket{1} + \ket{v}\ket{v}\,.
\end{equation}
Bob performs $V$ in his (spatially) local laboratory, and therefore the CJ states for the gates $B$ and $B^\prime$ are given as:
\begin{equation}
\label{eq:unitariesBob}
\begin{array}{rl}
	\kket{\tilde{V}^\ast}^{B_IB_O} & = \left[I^{B_IB_I} \otimes (\tilde{V}^\ast)^{B_OB_I}\right]\kket{\one}^{B_IB_I}\,, \vphantom{\ds \int} \\
	\kket{\tilde{V}^\ast}^{B_I^\prime B_O^\prime} & = \left[I^{B_I^\prime B_I^\prime} \otimes (\tilde{V}^\ast)^{B_O^\prime B_I^\prime}\right]\kket{\one}^{B_I^\prime B_I^\prime}\,.\vphantom{\ds \int}
\end{array}
\end{equation}

The gates $S^i$ and $S^f$ act as beam splitters, i.e., they both perform the same Hadamard operation $H$, given as follows. The beam splitter input and output spaces consist of the Alice's and Bob's factor spaces. For the case of the Alice's input space, we have $S_{A_I} = \sspan \{ \ket{0}^{S_{A_I}}, \ket{1}^{S_{A_I}}, \ket{v}^{S_{A_I}}\}$, and analogously for the output space, as well as for Bob's factor spaces. The overall input and output beam splitter spaces are therefore defined as $S_I = S_{(AB)_I} = S_{A_I} \otimes S_{B_I}$ and $S_O = S_{(AB)_O} = S_{A_O} \otimes S_{B_O}$. Finally, the unitary matrix associated to gate $S$ representing the action of the balanced Hadamard beam splitter is given by:
\begin{equation} \label{eq:hadamard_operator}
\begin{array}{rl}
  H^{S_OS_I} = & \ds\frac{1}{\sqrt 2} \left(\ket{0}^{S_{A_O}}\ket{v}^{S_{B_O}} + \ket{v}^{S_{A_O}}\ket{0}^{S_{B_O}}\right) \bra{0}^{S_{A_I}}\bra{v}^{S_{B_I}} \\
  & + \ds\frac{1}{\sqrt 2} \left(\ket{1}^{S_{A_O}}\ket{v}^{S_{B_O}} + \ket{v}^{S_{A_O}}\ket{1}^{S_{B_O}}\right) \bra{1}^{S_{A_I}}\bra{v}^{S_{B_I}} \\
  & + \ds\frac{1}{\sqrt 2} \left(\ket{0}^{S_{A_O}}\ket{v}^{S_{B_O}} - \ket{v}^{S_{A_O}}\ket{0}^{S_{B_O}}\right) \bra{v}^{S_{A_I}}\bra{0}^{S_{B_I}} \\
  & + \ds\frac{1}{\sqrt 2} \left(\ket{1}^{S_{A_O}}\ket{v}^{S_{B_O}} - \ket{v}^{S_{A_O}}\ket{1}^{S_{B_O}}\right) \bra{v}^{S_{A_I}}\bra{1}^{S_{B_I}} \, .
\end{array}
\end{equation}
The beam splitter acts such that the system coming from the Alice's side comes into an equal superposition of the two output spatial modes coming to Alice and Bob, with zero relative phase, while the system coming from the Bob's side (blue line) comes into an equal superposition of the two output spatial modes with relative phase $\pi$. Thus, in the output space the correlation between the Alice's and Bob's vacuum state is the opposite as in the input case. The corresponding CJ state is then
\begin{equation} \label{eq:hadamard}
	\kket{H^\ast}^{S_IS_O}  = \left[I^{S_IS_I} \otimes (H^\ast)^{S_OS_I}\right]\kket{\one}^{S_IS_I} \, ,
\end{equation}
where the transport vector $\kket{\one}$ for the beam splitter, when projected to a single-particle subspace, is given by
\begin{equation}
\kket{\one} = \ket{0v}\ket{0v} + \ket{1v}\ket{1v} + \ket{v0}\ket{v0} + \ket{v1}\ket{v1}\,.
\end{equation}
Note that the full transport vector contains nine terms instead of the above four, but for the purpose of this paper, we do not need those five additional terms.

The process vector encodes the wires between the gates, and it is being constructed by taking the tensor product over appropriate transport vectors $\kket{\one}$ for Alice's and Bob's qutrits, see equations (\ref{eq:TransportVectorsDef}) and (\ref{eq:TransportVectorsRelation}), such that each transport vector corresponds to one wire in the circuit diagram, connecting the output of the source gate to the input of the target gate. The process vector is thus given as:
\begin{equation} \label{eq:TotalProcessMatrix}
\begin{array}{lcl}
\kket{W_{\text{4-event}}} & = & \ds \underbrace{ \kket{\one}^{P_{A_O}S^i_{A_I}} \kket{\one}^{P_{B_O}S^i_{B_I}} }_{\text{initial}}
\underbrace{ \kket{\one}^{S^i_{A_O}A_I} \kket{\one}^{A_OB^\prime_I} \kket{\one}^{B^\prime_O S^f_{B_I}} }_{\text{blue}} \\
 & & \ds
\underbrace{ \kket{\one}^{S^i_{B_O}B_I} \kket{\one}^{B_OA^\prime_I} \kket{\one}^{A^\prime_O S^f_{A_I}} }_{\text{red}}
\underbrace{ \kket{\one}^{S^f_{A_O} T_{A_I}} \kket{\one}^{S^f_{B_O} T_{B_I}} }_{\text{final}} \,. \\
\end{array}
\end{equation}

One can now evaluate the probability distribution
\begin{equation} \label{eq:probability_from_amplitude}
p(\alpha,\beta) = \left\| \cM(\alpha,\beta) \vphantom{A^A_A} \right\|^2\,,
\end{equation}
where the probability amplitude $\cM(\alpha,\beta)$ is constructed by acting with the tensor product of all gate operations (\ref{eq:preparation_A}), (\ref{eq:hadamard}), (\ref{eq:unitaries}), (\ref{eq:unitariesBob}), (\ref{eq:hadamard}) and (\ref{eq:measurement_1}), on the process vector (\ref{eq:TotalProcessMatrix}). Since each of the gate operations acts in its own part of the total Hilbert space, the order of application of these operations is immaterial, and we are free to choose the most convenient one.

To see what happens when the operations (\ref{eq:preparation_A}) of the preparation gates act on the process vector, let us evaluate the action of $\kket{P_A^\ast}^{P_{A_I}P_{A_O}}$ on $\kket{\one}^{P_{A_O}S^i_{A_I}}$:
\begin{equation}
\bbra{P_A^\ast}^{P_{A_I}P_{A_O}} \kket{\one}^{P_{A_O}S^i_{A_I}} = \bra{\Psi^\ast}^{P_{A_O}} \sum_{k=0}^2 \ket{k}^{P_{A_O}} \ket{k}^{S^i_{A_I}} = \sum_{k=0}^2 \Big(\bracket{\Psi}{k}\Big)^\ast \ket{k}^{S^i_{A_I}} = \ket{\Psi}^{S^i_{A_I}}\,.
\end{equation}
An analogous calculation can be performed for $\kket{P_B^\ast}^{P_{B_I}P_{B_O}}$, so the action of both preparation operations (\ref{eq:preparation_A}) on the process vector (\ref{eq:TotalProcessMatrix}) evaluates to:
\begin{equation}
\begin{array}{l}
\ds \left( \bbra{P_A^\ast}^{P_{A_I}P_{A_O}} \otimes \bbra{P_B^\ast}^{P_{B_I}P_{B_O}} \right) \kket{W_{\text{4-event}}} = \vphantom{\ds\frac{1}{\sqrt{2}}} \\
\hphantom{mmmm} \ds 
\ket{\Psi}^{S^i_{A_I}} \ket{v}^{S^i_{B_I}}  
\underbrace{ \kket{\one}^{S^i_{A_O}A_I} \kket{\one}^{A_OB^\prime_I} \kket{\one}^{B^\prime_O S^f_{B_I}} }_{\text{blue}} \\
\hphantom{mmmm} \ds 
\underbrace{ \kket{\one}^{S^i_{B_O}B_I} \kket{\one}^{B_OA^\prime_I} \kket{\one}^{A^\prime_O S^f_{A_I}} }_{\text{red}}
\underbrace{ \kket{\one}^{S^f_{A_O} T_{A_I}} \kket{\one}^{S^f_{B_O} T_{B_I}}  }_{\text{final}}
 \,. \\
\end{array}
\end{equation}
Next one acts with the initial Hadamard operation (\ref{eq:hadamard}) on this process vector, transforming it into
\begin{equation}
\begin{array}{l}
\ds \left( \bbra{P_A^\ast}^{P_{A_I}P_{A_O}} \otimes \bbra{P_B^\ast}^{P_{B_I}P_{B_O}} \otimes \bbra{S^\ast}^{S^i_{(AB)_I} S^i_{(AB)_O}} \right) \kket{W_{\text{4-event}}} = \vphantom{\ds\frac{1}{\sqrt{2}}} \\
\hphantom{mmmm} \ds 
\frac{1}{\sqrt{2}} \left( \ket{\Psi}^{A_I} \ket{v}^{B_I} + \ket{v}^{A_I} \ket{\Psi}^{B_I} \right)
\underbrace{ \kket{\one}^{A_OB^\prime_I} \kket{\one}^{B^\prime_O S^f_{B_I}} }_{\text{blue}} \\
\hphantom{mmmm} \ds 
\underbrace{ \kket{\one}^{B_OA^\prime_I} \kket{\one}^{A^\prime_O S^f_{A_I}} }_{\text{red}}
\underbrace{ \kket{\one}^{S^f_{A_O} T_{A_I}} \kket{\one}^{S^f_{B_O} T_{B_I}} }_{\text{final}} \equiv \kket{W_{{QS}_4}}
 \,. \\
\end{array}
\end{equation}
The resulting process vector is the outcome of the action of the composite gate $\mathcal{I}$ on (\ref{eq:TotalProcessMatrix}), before the actions of Alice and Bob (note that often in the literature this is taken as the initial process vector in the analysis):
\begin{equation}
\label{eq:procmat_4_full}
\kket{W_{{QS}_4}} = \frac{1}{\sqrt 2} \left( \ket{\Psi}^{A_I}\ket{v}^{B_I} + \ket{v}^{A_I}\ket{\Psi}^{B_I} \right) \kket{\one}^{A_OB_I^\prime}\kket{\one}^{B_OA_I^\prime}\kket{\one}^{(A_O^\prime B_O^\prime)S_{(AB)_I}}\kket{\one}^{S_{(AB)_O}T_{(AB)_I}} \, .
\end{equation}
Continuing the computation, the action of the remaining gate operations (\ref{eq:unitaries}), (\ref{eq:unitariesBob}), (\ref{eq:measurement_1}) and (\ref{eq:hadamard}) on the process vector (\ref{eq:procmat_4_full}) gives us the probability amplitude,
$$
\cM(\alpha,\beta) \equiv \hphantom{mmmmmmmmmmmmmmmmmmmmmmmmmmmmmmmmmmmmmmmmm}
$$
\begin{equation}
\left( \bbra{\tilde{U}^\ast}^{A_IA_O} \otimes \bbra{\tilde{U}^\ast}^{A_I^\prime A_O^\prime} \otimes \bbra{\tilde{V}^\ast}^{B_IB_O} \otimes \bbra{\tilde{V}^\ast}^{B_I^\prime B_O^\prime}
 \otimes \bbra{H^\ast}^{S_IS_O} \otimes \bbra{T_{\alpha}^\ast}^{T_{A_I}T_{A_O}} \otimes \bbra{T_{\beta}^\ast}^{T_{B_I}T_{B_O}} \right) \kket{W_{{QS}_4}}\,.
\end{equation}
Let us now calculate the action of $\bbra{\tilde{U}^\ast}^{A_IA_O}$ on (\ref{eq:procmat_4_full}):
\begin{equation}
\bbra{\tilde{U}^\ast}^{A_IA_O} \kket{W_{{QS}_4}} = \bbra{\one}^{A_IA_I} \left[I^{A_IA_I} \otimes (\tilde{U}^T)^{A_OA_I}\right] \kket{W_{{QS}_4}} \,.
\end{equation}
Looking at the structure of the process vector, one sees that the resulting new process vector will have the form
\begin{equation} \label{eq:unevaluated_process_vector}
 \bbra{\tilde{U}^\ast}^{A_IA_O} \kket{W_{{QS}_4}} = \frac{1}{\sqrt 2} \left( \ket{X}^{B_I^\prime}\ket{v}^{B_I} + \ket{Y}^{B_I^\prime}\ket{\Psi}^{B_I} \right) \kket{\one}^{B_OA_I^\prime}\kket{\one}^{(A_O^\prime B_O^\prime)S_{(AB)_I}}\kket{\one}^{S_{(AB)_O}T_{(AB)_I}} \,,
\end{equation}
where $\ket{X}^{B_I^\prime}$ and $\ket{Y}^{B_I^\prime}$ are shorthands for the expressions
\begin{equation}
\ket{X}^{B_I^\prime} \equiv \bbra{\one}^{A_IA_I} \left[I^{A_IA_I} \otimes (\tilde{U}^T)^{A_OA_I}\right] \ket{\Psi}^{A_I} \kket{\one}^{A_OB_I^\prime}
\end{equation}
and
\begin{equation}
\ket{Y}^{B_I^\prime} \equiv \bbra{\one}^{A_IA_I} \left[I^{A_IA_I} \otimes (\tilde{U}^T)^{A_OA_I}\right] \ket{v}^{A_I} \kket{\one}^{A_OB_I^\prime}\,,
\end{equation}
which need to be evaluated. The explicit computation of the first expression goes as follows:
\begin{equation} \label{eq:computation_of_X}
\begin{array}{lcl}
\ket{X}^{B_I^\prime} & = & \ds \bbra{\one}^{A_IA_I} \left[I^{A_IA_I} \otimes (\tilde{U}^T)^{A_OA_I}\right] \ket{\Psi}^{A_I} \kket{\one}^{A_OB_I^\prime} \vphantom{\ds\sum_{k=0}^2} \\
& = & \ds \sum_{k=0}^2 \bra{k}^{A_I} \bra{k}^{A_I} \left[I^{A_IA_I} \otimes (\tilde{U}^T)^{A_OA_I}\right] \ket{\Psi}^{A_I} \sum_{m=0}^2 \ket{m}^{A_O} \ket{m}^{B_I^\prime} \\
& = & \ds \sum_{m=0}^2\left[ \sum_{k=0}^2  \left( \bra{k}^{A_I} I^{A_IA_I} \ket{\Psi}^{A_I} \right) \left( \bra{k}^{A_I} (\tilde{U}^T)^{A_OA_I} \ket{m}^{A_O} \right) \right] \ket{m}^{B_I^\prime} \\
& = & \ds \sum_{m=0}^2\left[ \sum_{k=0}^2 \bracket{k}{\Psi}^{A_I} \; \bra{m}^{A_O} \tilde{U}^{A_OA_I} \ket{k}^{A_I}  \right] \ket{m}^{B_I^\prime} \\
& = & \ds \sum_{m=0}^2\left[ \bra{m}^{A_O} \tilde{U}^{A_OA_I} \ket{\Psi}^{A_I} \right] \ket{m}^{B_I^\prime}\,. \\
\end{array}
\end{equation}
Using (\ref{eq:u_tilde_def}), the coefficient in the brackets can be evaluated as
\begin{equation}
\bra{m}^{A_O} \tilde{U}^{A_OA_I} \ket{\Psi}^{A_I} = \bra{m}^{A_O} \left( U^{A_OA_I}P_{01}^{A_IA_I} + I^{A_OA_I}P_v^{A_IA_I} \right) \ket{\Psi}^{A_I} = \bra{m} U \ket{\Psi}\,,
\end{equation}
since $P_{01}^{A_IA_I} \ket{\Psi}^{A_I} = \ket{\Psi}^{A_I}$ and $P_v^{A_IA_I} \ket{\Psi}^{A_I} = 0$. Thus, we have
\begin{equation} \label{eq:result_for_X}
\ket{X}^{B_I^\prime} = \sum_{m=0}^2 \bra{m} U \ket{\Psi}\; \ket{m}^{B_I^\prime} = U\ket{\Psi}^{B_I^\prime} \equiv \ket{U\Psi}^{B_I^\prime}\,.
\end{equation}
The computation of $\ket{Y}^{B_I^\prime}$ proceeds in an analogous way to (\ref{eq:computation_of_X}), and the result is
\begin{equation}
\ket{Y}^{B_I^\prime} = \sum_{m=0}^2\left[ \bra{m}^{A_O} \tilde{U}^{A_OA_I} \ket{v}^{A_I} \right] \ket{m}^{B_I^\prime}\,.
\end{equation}
Again, using  (\ref{eq:u_tilde_def}), the coefficient in the brackets can be evaluated as
\begin{equation}
\bra{m}^{A_O} \tilde{U}^{A_OA_I} \ket{v}^{A_I} = \bra{m}^{A_O} \left( U^{A_OA_I}P_{01}^{A_IA_I} + I^{A_OA_I}P_v^{A_IA_I} \right) \ket{v}^{A_I} = \bracket{m}{v} = \delta_{mv}\,,
\end{equation}
since $P_{01}^{A_IA_I} \ket{v}^{A_I} = 0$ and $P_v^{A_IA_I} \ket{v}^{A_I} = \ket{v}^{A_I}$. Thus, we have
\begin{equation} \label{eq:result_for_Y}
\ket{Y}^{B_I^\prime} = \sum_{m=0}^2 \delta_{mv}\; \ket{m}^{B_I^\prime} = \ket{v}^{B_I^\prime}\,.
\end{equation}
Finally, substituting (\ref{eq:result_for_X}) and (\ref{eq:result_for_Y}) back into (\ref{eq:unevaluated_process_vector}), we obtain:
\begin{equation} \label{eq:partially_evaluated_process_vector}
\bbra{\tilde{U}^\ast}^{A_IA_O} \kket{W_{{QS}_4}} = \frac{1}{\sqrt 2} \left( \ket{U\Psi}^{B_I^\prime}\ket{v}^{B_I} + \ket{v}^{B_I^\prime}\ket{\Psi}^{B_I} \right) \kket{\one}^{B_OA_I^\prime}\kket{\one}^{(A_O^\prime B_O^\prime)S_{(AB)_I}}\kket{\one}^{S_{(AB)_O}T_{(AB)_I}} \,. 
\end{equation}

One should note, comparing (\ref{eq:partially_evaluated_process_vector}) with (\ref{eq:procmat_4_full}), that the action of the gate $A$ operation onto the process vector effectively performs the following transformation,
\begin{equation}
\ket{\Psi}^{A_I} \to \ket{U\Psi}^{A_O} \to \ket{U\Psi}^{B_I^\prime} \,, \qquad \ket{v}^{A_I} \to \ket{v}^{A_O} \to \ket{v}^{B_I^\prime} \,,
\end{equation}
where the transport vector $\kket{\one}^{A_OB_I^\prime}$ has been utilised for ``transporting'' the state from the output $A_O$ of gate $A$ to the input $B_I^\prime$ of the gate $B^\prime$, in line with the spacetime diagram. This scheme repeats itself with the action of all remaining gate operations on (\ref{eq:partially_evaluated_process_vector}). In particular, the subsequent action of the gate $B$ operation gives:
$$
\left( \bbra{\tilde{V}^\ast}^{B_IB_O} \otimes \bbra{\tilde{U}^\ast}^{A_IA_O} \right) \kket{W_{{QS}_4}} = \hphantom{mmmmmmmmmmmmmmmmmmmmmmm}
$$
\begin{equation}
\frac{1}{\sqrt 2} \left( \ket{U\Psi}^{B_I^\prime}\ket{v}^{A_I^\prime} + \ket{v}^{B_I^\prime}\ket{V\Psi}^{A_I^\prime} \right) \kket{\one}^{(A_O^\prime B_O^\prime)S_{(AB)_I}}\kket{\one}^{S_{(AB)_O}T_{(AB)_I}} \,,
\end{equation}
which can also be verified with an explicit calculation similar to the above. Continuing on, the operations at the gates $A^\prime$ and $B^\prime$ give:
\begin{equation} \label{eq:temp_process_vector_one}
\begin{array}{l}
\ds \left( \bbra{\tilde{V}^\ast}^{B_I^\prime B_O^\prime} \otimes \bbra{\tilde{U}^\ast}^{A_I^\prime A_O^\prime} \otimes \bbra{\tilde{V}^\ast}^{B_IB_O} \otimes \bbra{\tilde{U}^\ast}^{A_IA_O} \right) \kket{W_{{QS}_4}} = \vphantom{\ds\frac{1}{\sqrt{2}}} \\
\hphantom{mmmm} \ds \frac{1}{\sqrt 2} \left( \ket{VU\Psi}^{S_{B_I}}\ket{v}^{S_{A_I}} + \ket{v}^{S_{B_I}}\ket{UV\Psi}^{S_{A_I}} \right) \kket{\one}^{S_{(AB)_O}T_{(AB)_I}} \,. \\
\end{array}
\end{equation}
Next, the action of the beam splitter at the gate $S_f$ gives
\begin{equation} \label{eq:temp_process_vector_two}
\begin{array}{l}
\ds \left( \bbra{\tilde{H}^\ast}^{S_IS_O} \otimes \bbra{\tilde{V}^\ast}^{B_I^\prime B_O^\prime} \otimes \bbra{\tilde{U}^\ast}^{A_I^\prime A_O^\prime} \otimes \bbra{\tilde{V}^\ast}^{B_IB_O} \otimes \bbra{\tilde{U}^\ast}^{A_IA_O} \right) \kket{W_{{QS}_4}} = \vphantom{\ds\frac{1}{\sqrt{2}}} \\
\hphantom{mmmm} \ds \frac{1}{2} \sum_{i=0}^1 \left( \bra{i}\akomut{U}{V}\ket{\Psi}\; \ket{i}^{T_{A_I}}\ket{v}^{T_{B_I}} + \bra{i}\komut{U}{V} \ket{\Psi}\; \ket{v}^{T_{A_I}}\ket{i}^{T_{B_I}} \right) \,. \\
\end{array}
\end{equation}
Finally, the action of the operations of the target gates $T_A$ and $T_B$ gives us the probability amplitude as a function of the measurement outcomes $\alpha$ and $\beta$,
$$
\cM(\alpha,\beta) \equiv \hphantom{mmmmmmmmmmmmmmmmmmmmmmmmmm}
$$
$$
\left( \bbra{\tilde{U}^\ast}^{A_IA_O} \otimes \bbra{\tilde{U}^\ast}^{A_I^\prime A_O^\prime} \otimes \bbra{\tilde{V}^\ast}^{B_IB_O} \otimes \bbra{\tilde{V}^\ast}^{B_I^\prime B_O^\prime} \otimes \bbra{H^\ast}^{S_IS_O} \otimes \bbra{T_{\alpha}^\ast}^{T_{A_I}T_{A_O}} \otimes \bbra{T_{\beta}^\ast}^{T_{B_I}T_{B_O}} \right) \kket{W_{{QS}_4}}
$$
\begin{equation}
 = \frac{1}{2} \Big[ \delta_{\beta v} \bra{\alpha} \akomut{U}{V} \ket{\Psi} + \delta_{\alpha v} \bra{\beta} \komut{U}{V} \ket{\Psi} \Big] \,.
\end{equation}
At this point we can employ (\ref{eq:probability_from_amplitude}) and calculate the probability distribution,
\begin{equation} \label{eq:final_probability_distribution}
p(\alpha,\beta) = \frac{1}{4} \Big[ \delta_{\beta v} \left|  \bra{\alpha} \akomut{U}{V} \ket{\Psi} \vphantom{A^A} \right|^2 + \delta_{\alpha v} \left| \bra{\beta} \komut{U}{V} \ket{\Psi} \vphantom{A^A} \right|^2 \Big] \,, \\
\end{equation}
where we have used the fact that the vacuum state $\ket{v}$ is orthogonal to both $\akomut{U}{V} \ket{\Psi}$ and $\komut{U}{V} \ket{\Psi}$. In particular, we see that for $i\in\{0,1\}$ we have
\begin{equation} \label{eq:nonzero_probability_values}
p(i,v) = \frac{1}{4} \left| \bra{i} \akomut{U}{V} \ket{\Psi} \vphantom{A^A} \right|^2 \,, \qquad
p(v,i) = \frac{1}{4} \left| \bra{i} \komut{U}{V} \ket{\Psi} \vphantom{A^A} \right|^2 \,,
\end{equation}
while all other choices of $\alpha$ and $\beta$ give $p(\alpha,\beta)=0$. The total probability that Alice will detect a particle is given by the marginal
\begin{equation}
p_A = \sum_{i=1}^2 p(i,v) = \frac{1}{2} \left( 1 + \Re \bra{\Psi}U^\dag V^\dag UV \ket{\Psi} \vphantom{A^A} \right)\,,
\end{equation}
while the corresponding total probability that Bob will detect a particle is
\begin{equation}
p_B = \sum_{i=1}^2 p(v,i) = \frac{1}{2} \left( 1 - \Re \bra{\Psi}U^\dag V^\dag UV \ket{\Psi} \vphantom{A^A} \right)\,.
\end{equation}
As a final point, we can verify that the probability distribution is correctly normalised. Using the fact that the only nonzero values for the probability are given in (\ref{eq:nonzero_probability_values}), we have
\begin{equation}
p_{\text{total}} = \sum_{\alpha=0}^2 \sum_{\beta=0}^2 p(\alpha,\beta) = \underbrace{p(0,v)+p(1,v)}_{p_A} + \underbrace{p(v,0)+p(v,1)}_{p_B} = 1\,,
\end{equation}
as expected.

Instead of recombining the particle on the second beam splitter, one can consider the case in which the final gate $\mathcal{F}$ consists only of local measurements performed onto a particle in the Alice's and Bob's paths. In this case, the final gate is equivalent to two target gates $T_A$ and $T_B$. Given that all the processes considered are pure, the corresponding process vector for the 4-event quantum switch implementation is given as (in order to compare the current with the previous works, we present the process that starts after $\mathcal{I}$, as was done in, say,~\cite{ara:bra:cos:fei:gia:bru:15}):
\begin{equation}
\label	{eq:procmat_4}
\kket{W_{{QS}_4}} = \frac{1}{\sqrt 2} \left( \ket{\Psi}^{A_I}\ket{v}^{B_I} + \ket{v}^{A_I}\ket{\Psi}^{B_I} \right) \kket{\one}^{A_OB_I^\prime}\kket{\one}^{B_OA_I^\prime}\kket{\one}^{A_O^\prime T_{A_I}}\kket{\one}^{B_O^\prime T_{B_I}}.
\end{equation}

\section{\label{sec:AppThreeEvent}3-event process vector}

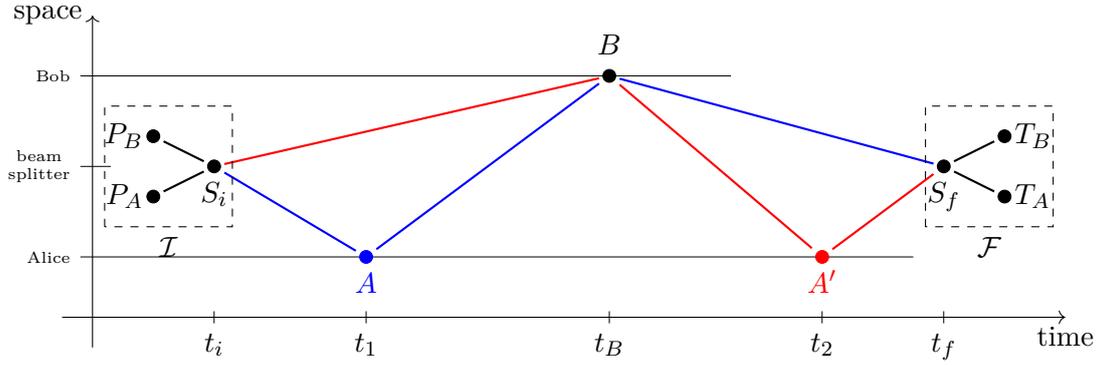
\begin{figure}[!ht]
\begin{center}
\begin{tikzpicture}[x={(0cm,1cm)},y={(1cm,0cm)},scale=0.8]
\draw[->] (3.5,-1.5) -- (9,-1.5);
\node[anchor=east] at (9,-1.5) {space};

\draw[->] (4,-2) -- (4,14.5);
\node[anchor=north] at (4,14.5) {time};
\draw[very thin] (3.9,0.5) -- (4.1,0.5);
\node[anchor=north] at (3.9,0.5) {$t_i$};
\draw[very thin] (3.9,3) -- (4.1,3);
\node[anchor=north] at (3.9,3) {$t_1$};
\draw[very thin] (3.9,7) -- (4.1,7);
\node[anchor=north] at (3.9,7) {$t_B$};
\draw[very thin] (3.9,10.5) -- (4.1,10.5);
\node[anchor=north] at (3.9,10.5) {$t_2$};
\draw[very thin] (3.9,12.5) -- (4.1,12.5);
\node[anchor=north] at (3.9,12.5) {$t_f$};

\draw[very thin] (5,-1.7) -- (5,12);
\node[anchor=east] at (5,-1.7) {\tiny Alice};

\draw[very thin] (8,-1.7) -- (8,9);
\node[anchor=east] at (8,-1.7) {\tiny Bob};

\node at (6.5,0.5) (splitterSi) {};
\node at (6.5,12.5) (splitterSf) {};
\draw[very thin] (6.5,-1.7) -- (6.5,-1.2);
\node[anchor=east] at (6.5,-1.7) {$\substack{ \text{\tiny beam} \\ \text{\tiny splitter} }$};
\filldraw[black] (splitterSi) circle (3pt) node[anchor=north] {$S_i\strut$};
\filldraw[black] (splitterSf) circle (3pt) node[anchor=north] {$S_f\strut$};

\node at (5,3) (bluegateA) {};
\node at (8,7) (gateB) {};
\filldraw[blue] (bluegateA) circle (3pt) node[anchor=north] {$A\strut$};
\filldraw[black] (gateB) circle (3pt) node[anchor=south] {$B\strut$};

\node at (5,10.5) (redgateAprime) {};
\filldraw[red] (redgateAprime) circle (3pt) node[anchor=north] {$A^\prime\strut$};

\draw[thick,blue] (splitterSi) -- (bluegateA) -- (gateB) -- (splitterSf);

\draw[thick,red] (splitterSi) -- (gateB) -- (redgateAprime) -- (splitterSf);


\node at (6,-0.5) (gatePa) {};
\node at (7,-0.5) (gatePb) {};
\filldraw[black] (gatePa) circle (3pt) node[anchor=east] {$P_A\strut$};
\filldraw[black] (gatePb) circle (3pt) node[anchor=east] {$P_B\strut$};
\draw[thick] (gatePa) -- (splitterSi);
\draw[thick] (gatePb) -- (splitterSi);
\draw[dashed] (5.5,-1.3) rectangle (7.5,0.8);
\node[anchor=north] at (5.5,-0.25) {$\mathcal{I}$};

\node at (6,13.5) (gateTa) {};
\node at (7,13.5) (gateTb) {};
\filldraw[black] (gateTa) circle (3pt) node[anchor=west] {$T_A\strut$};
\filldraw[black] (gateTb) circle (3pt) node[anchor=west] {$T_B\strut$};
\draw[thick] (splitterSf) -- (gateTa);
\draw[thick] (splitterSf) -- (gateTb);
\draw[dashed] (5.5,12.2) rectangle (7.5,14.3);
\node[anchor=north] at (5.5,13.25) {$\mathcal{F}$};

\end{tikzpicture}
\end{center}
\caption{\blue{Spacetime diagram of the $3$-event implementation of the quantum switch. The internal structures of the composite gates $\cI$ and $\cF$ are explicitly depicted.}}
\label{sl:petnaest}
\end{figure}

The detailed spacetime diagram for the $3$-event quantum switch is given below \blue{(see Figure~\ref{sl:petnaest})}.

The process vector for this case is obtained from~\eqref{eq:TotalProcessMatrix} by identifying the spacetime positions of the gates $B$ and $B^\prime$, yet keeping the corresponding Hilbert spaces in the mathematical description (i.e., keeping the dimensionality of the problem). Thus, the corresponding circuit is {\em identical} to the $4$-event circuit, and the process vector has the {\em identical} mathematical form as in the case of four gates. In order to emphasise the physical difference between the two cases, instead of $B_{I/O}$ and $B_{I/O}^\prime$, we write $B_{I_1/O_1}$ and $B_{I_2/O_2}$, respectively:
\begin{equation}
\label	{eq:procmat_3}
\begin{array}{lcl}
\kket{W_{\text{3-event}}} & = & \ds \underbrace{ \kket{\one}^{P_{A_O}S^i_{A_I}} \kket{\one}^{P_{B_O}S^i_{B_I}} }_{\text{initial}}
\underbrace{ \kket{\one}^{S^i_{A_O}A_I} \kket{\one}^{A_OB_{I_2}} \kket{\one}^{B_{O_2} S^f_{B_I}} }_{\text{blue}} \\
 & & \ds
\underbrace{ \kket{\one}^{S^i_{B_O}B_{I_1}} \kket{\one}^{B_{O_1}A^\prime_I} \kket{\one}^{A^\prime_O S^f_{A_I}} }_{\text{red}}
\underbrace{ \kket{\one}^{S^f_{A_O} T_{A_I}} \kket{\one}^{S^f_{B_O} T_{B_I}} }_{\text{final}} \,. \\
\end{array}
\end{equation}
The final probability distribution is identical to the one for the 4-event process, given by~\eqref{eq:final_probability_distribution}.

\section{\label{sec:AppTwoEvent}2-event}

In this Appendix we present process vectors for the two gravitational switches discussed in the main text. First, the process vector of the gravitational switch without recombination~\cite{zyc:cos:pik:bru:17}, is given by (since the ``control'' is now played by gravity, it is thus denoted as $G$, instead of $C$):
\begin{equation}
\label{eq:procmat_2_mag}
\kket{W_{{QS}_2}} = \frac{1}{\sqrt 2} \left( \ket{0}^G\ket{\Psi}^{A_I}\kket{\one}^{A_OB_I}\kket{\one}^{B_O T_{B_I}} + \ket{1}^G\ket{\Psi}^{B_I}\kket{\one}^{B_OA_I}\kket{\one}^{A_O T_{A_I}} \right)\kket{\one}^{G T_{G_I}}.
\end{equation}
It is then straightforward to obtain the process vector for the case of recombining only the gravity on the final beam splitter $S_f$ (a part of a bigger final gate $\mathcal{F}$), obtaining the analogue of~\eqref{eq:procmat_4_full}, while the particle is not being recombined. Note that in this case the introduction of the vacuum state is not necessary, as in each branch of superposition all gates are acting upon a particle.

In contrast to the above case, the process vector describing the gravitational 2-event quantum switch with the recombination of both gravity and the particle is given as follows:
\begin{equation}
\label{eq:procmat_2_rec}
\begin{array}{rl}
\kket{W_{{QS}_2}^{(r)}} = \ds\frac{1}{\sqrt 2} \!\!\!\! &  \left( \ket{0}^G\ket{\Psi}^{A_I}\kket{\one}^{A_OB_I}\kket{\one}^{B_O S_{P_I}} + \ket{1}^G\ket{\Psi}^{B_I}\kket{\one}^{B_OA_I}\kket{\one}^{A_O S_{P_I}} \right) \\ & \otimes \kket{\one}^{G S_{G_I}} \kket{\one}^{(S_{G_O}S_{P_O})(T_{G_I}T_{P_I})}.
\end{array}
\end{equation}
Here, $P$ stands for ``the particle'' (whose corresponding input space $S_{P_I}$ is isomorphic to the tensor product of Alice's and Bob's output spaces, $S_{P_I} \simeq A_O \otimes B_O$), and $S$ stands for a ``beam splitter'' (whose corresponding input space is $S_I = S_{G_I} \otimes S_{P_I}$, and analogously for the output space).

While defining the spaces $S_{G_{I/O}}$, $S_{P_{I/O}}$, $T_{G_I}$, $T_{P_I}$, and the vector $\kket{\one}^{(S_{G_O}S_{P_O})(T_{G_I}T_{P_O})}$ is straightforward, it is not so for the ``final'' recombination vector $\kket{U_{BS}^\ast}^{(S_{G_I}S_{P_I})(S_{G_O}S_{P_O})}$. Namely, note that in our gravitational switch {\em all} degrees of freedom, both gravitational and matter, are recombined by $U_{BS}$ such that, by acting on the beam splitter input entangled state,
\begin{equation}
\label{eq:entangled_input}
	\ket{\Psi_i}^{S_{G_I}S_{P_I}} =\frac{1}{\sqrt 2} \left( \ket{0}^{S_{G_I}} \otimes \left[ UV\ket{\Psi}^{S_{P_I}} \right] + \ket{1}^{S_{G_I}} \otimes \left[ VU\ket{\Psi}^{S_{P_I}} \right] \right),
\end{equation}
the overall output state is a product one, of the form
\begin{equation}
\label{eq:product_output}
	\ket{\Psi_o}^{S_{G_O}S_{P_O}} = \ket{\Gamma}^{S_{G_O}} \otimes \left( \alpha UV + \beta VU \right) \ket{\Psi}^{S_{P_O}},
\end{equation}
where $\ket{\Gamma}^{S_{G_O}}$ is some (not necessarily classical) state of gravity. The above evolution is, at least in principle, allowed by the quantum laws, which is all that we need to know regarding the action of $U_{BS}$ at this point. Its action on the rest of the overall Hilbert space is, for the purpose of our argument, irrelevant, and can thus be chosen arbitrarily. 

Finally, we would like to note that the same type of the 4-, or 3-event quantum switches in classical spacetimes can also be defined, resulting in the same type of the output state as~\eqref{eq:product_output}, with the gravity degree of freedom being replaced by any additional matter degree of freedom that plays the role of the control.

\section{\label{app:gravitacioni_svicevi}Various implementations of the gravitational switch}

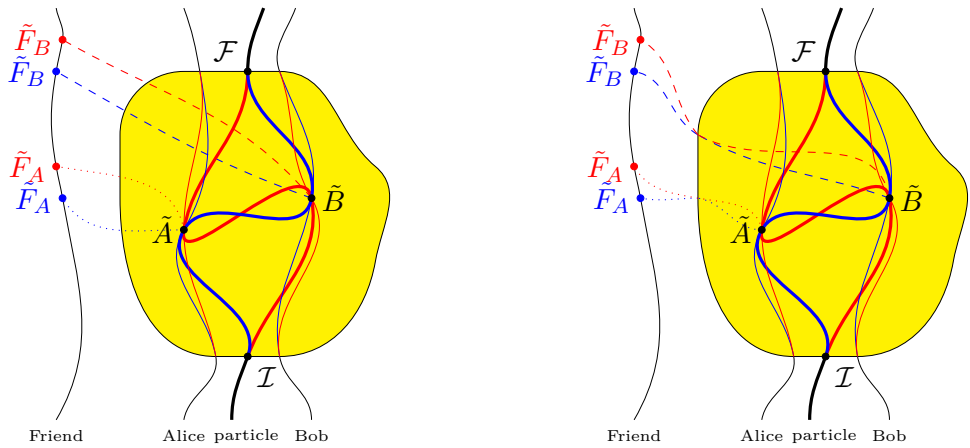
\begin{figure}[h!]
\begin{center}
\begin{tikzpicture}[scale=0.42]

\draw[very thin, fill=yellow] (5,2) to [out=0, in=180 ] (7,2) to [out=0, in=260] (10,5) to [out=80, in=320] (10,8) to [out=140, in=0] (7,11) to [out=180, in=0] (4,11) to [out=180, in=90] (2,9) to [out=270, in=90] (2,7) to [out=270, in=180] (5,2);

\draw[very thin] (0,0) to [out=60, in=280 ] (0.2,7) to [out=100, in=280 ] (0,8) to [out=100, in=260] (0,11) to [out=80, in=260] (0.2,12) to [out=80, in=280] (0,13);
\node at (0,-0.5) {\tiny Friend};

\draw[dashed, blue] (8,7) to [out=160, in=330] (0,11);
\draw[dashed, red] (8,7) to [out=130, in=330] (0.2,12);

\draw[dotted, blue] (4,6) to [out=180, in=300] (0.2,7);
\draw[dotted, red] (4,6) to [out=100, in=350] (0,8);

\node at (5.5,-0.5) {$\quad$\tiny particle};
\draw[very thick, black] (5.5,0) to [out=90 , in=250] (6,2);
\draw[very thick, black] (6,11) to [out=90 , in=250] (6.5,13);

\draw[very thick, red] (6,2) to [out=70, in=280] (8,7) to [out=100 , in=260] (4,6) to [out=80 , in=270] (6,11); 

\draw[very thick, blue] (6,2) to [out=70, in=240] (4,6) to [out=60 , in=260] (8,7) to [out=80 , in=270] (6,11); 

\node at (4,-0.5) {\tiny Alice};
\draw[very thin, black] (4,0) to [out=80, in=280] (5,2);
\draw[very thin, black] (4.5,11) to [out=100, in=290] (4,13);

\draw[very thin, blue] (5,2) to [out=100, in=240] (4,6) to [out=60, in=280] (4.5,11);

\draw[very thin, red] (5,2) to [out=100, in=270] (4,6) to [out=90, in=280] (4.5,11);

\node at (8,-0.5) {\tiny Bob};
\draw[very thin, black] (8,0) to [out=90, in=280] (7,2);
\draw[very thin, black] (7,11) to [out=110, in=250] (7.5,13);

\draw[very thin, blue] (7,2) to [out=100, in=260] (8,7) to [out=80, in=290] (7,11);

\draw[very thin, red] (7,2) to [out=100, in=300] (8,7) to [out=120, in=290] (7,11);

\filldraw[black] (4,6) circle (3pt) node[anchor=east] {$\tilde A$};
\filldraw[black] (8,7) circle (3pt) node[anchor=west] {$\tilde B$};
\filldraw[blue] (0.2,7) circle (3pt) node[anchor=east] {$\tilde{F}_A$};
\filldraw[red] (0,8) circle (3pt) node[anchor=east] {$\tilde{F}_A$};
\filldraw[blue] (0,11) circle (3pt) node[anchor=east] {$\tilde{F}_B$};
\filldraw[red] (0.2,12) circle (3pt) node[anchor=east] {$\tilde{F}_B$};
\filldraw[black] (6,2) circle (3pt) node[anchor=north west] {$\mathcal{I}$};
\filldraw[black] (6,11) circle (3pt) node[anchor=south east] {$\mathcal{F}$};

\end{tikzpicture}
\hspace{2cm}
\begin{tikzpicture}[scale=0.42]

\draw[very thin, fill=yellow] (5,2) to [out=0, in=180 ] (7,2) to [out=0, in=260] (10,5) to [out=80, in=320] (10,8) to [out=140, in=0] (7,11) to [out=180, in=0] (4,11) to [out=180, in=90] (2,9) to [out=270, in=90] (2,7) to [out=270, in=180] (5,2);

\draw[very thin] (0,0) to [out=60, in=280 ] (0.2,7) to [out=100, in=280 ] (0,8) to [out=100, in=260] (0,11) to [out=80, in=260] (0.2,12) to [out=80, in=280] (0,13);
\node at (0,-0.5) {\tiny Friend};

\draw[dashed, blue] (8,7) to [out=160, in=330] (2,9);
\draw[dashed, blue] (2,9) to [out=150, in=350] (0,11);
\draw[dashed, red] (8,7) to [out=100, in=330] (2,9);
\draw[dashed, red] (2,9) to [out=130, in=330] (0.2,12);

\draw[dotted, blue] (4,6) to [out=180, in=350] (2,7);
\draw[dotted, blue] (2,7) to [out=180, in=350] (0.2,7);
\draw[dotted, red] (4,6) to [out=100, in=350] (2,7);
\draw[dotted, red] (2,7) to [out=170, in=300] (0,8);

\node at (5.5,-0.5) {$\quad$\tiny particle};
\draw[very thick, black] (5.5,0) to [out=90 , in=250] (6,2);
\draw[very thick, black] (6,11) to [out=90 , in=250] (6.5,13);

\draw[very thick, red] (6,2) to [out=70, in=280] (8,7) to [out=100 , in=260] (4,6) to [out=80 , in=270] (6,11); 

\draw[very thick, blue] (6,2) to [out=70, in=240] (4,6) to [out=60 , in=260] (8,7) to [out=80 , in=270] (6,11); 

\node at (4,-0.5) {\tiny Alice};
\draw[very thin, black] (4,0) to [out=80, in=280] (5,2);
\draw[very thin, black] (4.5,11) to [out=100, in=290] (4,13);

\draw[very thin, blue] (5,2) to [out=100, in=240] (4,6) to [out=60, in=280] (4.5,11);

\draw[very thin, red] (5,2) to [out=100, in=270] (4,6) to [out=90, in=280] (4.5,11);

\node at (8,-0.5) {\tiny Bob};
\draw[very thin, black] (8,0) to [out=90, in=280] (7,2);
\draw[very thin, black] (7,11) to [out=110, in=250] (7.5,13);

\draw[very thin, blue] (7,2) to [out=100, in=260] (8,7) to [out=80, in=290] (7,11);

\draw[very thin, red] (7,2) to [out=100, in=300] (8,7) to [out=120, in=290] (7,11);

\filldraw[black] (4,6) circle (3pt) node[anchor=east] {$\tilde A$};
\filldraw[black] (8,7) circle (3pt) node[anchor=west] {$\tilde B$};
\filldraw[blue] (0.2,7) circle (3pt) node[anchor=east] {$\tilde{F}_A$};
\filldraw[red] (0,8) circle (3pt) node[anchor=east] {$\tilde{F}_A$};
\filldraw[blue] (0,11) circle (3pt) node[anchor=east] {$\tilde{F}_B$};
\filldraw[red] (0.2,12) circle (3pt) node[anchor=east] {$\tilde{F}_B$};
\filldraw[black] (6,2) circle (3pt) node[anchor=north west] {$\mathcal{I}$};
\filldraw[black] (6,11) circle (3pt) node[anchor=south east] {$\mathcal{F}$};

\end{tikzpicture}
\end{center}
\caption{\blue{The spacetime diagrams of a $2$-event gravitational switch implementations, with Friend's measurements, which fail to distinguish them from the optical implementation of the quantum switch.}}
\label{sl:sesnaest}
\end{figure}

In this Appendix we present a few representative additional constructions of the gravitational quantum switch. First, we start with a $2$-event switch for which the requirement (i) from Subsection \ref{sec:otherswitches} is not satisfied. It is obvious from the diagram on the left that each of the photon's superposed trajectories fail to meet at the boundary region, violating requirement (i), \blue{see the left diagram of Figure~\ref{sl:sesnaest}.} Next, we proceed with the $2$-event implementation for which requirement (i) is satisfied, but the requirement (ii) is not, since the superposed trajectories fail to recombine. This is depicted on the \blue{right diagram of Figure~\ref{sl:sesnaest}}.

\begin{figure}[!ht]
\begin{center}
\begin{tikzpicture}[scale=0.42]

\draw[very thin, fill=yellow] (5,2) to [out=0, in=180 ] (7,2) to [out=0, in=260] (10,5) to [out=80, in=320] (10,8) to [out=140, in=0] (7,11) to [out=180, in=0] (4,11) to [out=180, in=90] (2,9) to [out=270, in=90] (2,7) to [out=270, in=180] (5,2);

\draw[very thin] (0,0) to [out=60, in=280 ] (0,8) to [out=100, in=260] (0,11) to [out=80, in=280] (0,13);
\node at (0,-0.5) {\tiny Friend};

\draw[dashed, blue] (8,8) to [out=160, in=330] (2,9);
\draw[dashed, red] (7.5,6) to [out=100, in=330] (2,9);
\draw[dashed, black] (2,9) to [out=150, in=350] (0,11);

\draw[dotted, blue] (4.5,5) to [out=180, in=350] (2,7);
\draw[dotted, red] (4,7) to [out=100, in=350] (2,7);
\draw[dotted, black] (2,7) to [out=170, in=300] (0,8);

\node at (5.5,-0.5) {$\quad$\tiny particle};
\draw[very thick, black] (5.5,0) to [out=90 , in=250] (6,2);
\draw[very thick, black] (6,11) to [out=90 , in=250] (6.5,13);

\draw[very thick, red] (6,2) to [out=70, in=280] (7.5,6) to [out=100 , in=260] (4,7) to [out=80 , in=270] (6,11); 

\draw[very thick, blue] (6,2) to [out=70, in=240] (4.5,5) to [out=60 , in=260] (8,8) to [out=80 , in=270] (6,11); 

\node at (4,-0.5) {\tiny Alice};
\draw[very thin, black] (4,0) to [out=80, in=300] (5,2);
\draw[very thin, black] (4.5,11) to [out=100, in=290] (4,13);

\draw[very thin, blue] (5,2) to [out=110, in=240] (4.5,5) to [out=60, in=280] (4.5,11);

\draw[very thin, red] (5,2) to [out=130, in=270] (4,7) to [out=90, in=280] (4.5,11);

\node at (8,-0.5) {\tiny Bob};
\draw[very thin, black] (8,0) to [out=90, in=240] (7,2);
\draw[very thin, black] (7,11) to [out=110, in=250] (7.5,13);

\draw[very thin, blue] (7,2) to [out=60, in=270] (8,8) to [out=80, in=290] (7,11);

\draw[very thin, red] (7,2) to [out=70, in=280] (7.5,6) to [out=100, in=290] (7,11);

\filldraw[blue] (4.5,5) circle (3pt) node[anchor=west] {$A$};
\filldraw[red] (4,7) circle (3pt) node[anchor=east] {$A^\prime$};
\filldraw[red] (7.5,6) circle (3pt) node[anchor=east] {$B$};
\filldraw[blue] (8,8) circle (3pt) node[anchor=west] {$B^\prime$};
\filldraw[black] (0,8) circle (3pt) node[anchor=east] {$F_{AA^\prime}$};
\filldraw[black] (0,11) circle (3pt) node[anchor=east] {$F_{BB^\prime}$};
\filldraw[black] (6,2) circle (3pt) node[anchor=north west] {$\mathcal{I}$};
\filldraw[black] (6,11) circle (3pt) node[anchor=south east] {$\mathcal{F}$};

\end{tikzpicture}
\end{center}
\caption{\blue{The spacetime diagram of a $4$-event gravitational switch implementation, with Friend's measurement, which fails to distinguish it from the $2$-event implementation of the gravitational switch.}}
\label{sl:sedamnaest}
\end{figure}
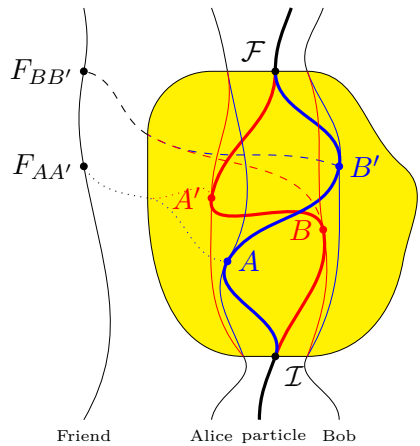

Finally, we present a $4$-event implementation for which {\em both} requirements (i) and (ii) are satisfied \blue{(see Figure~\ref{sl:sedamnaest})}.

Of course, other variations are possible as well.

\end{document}